\documentclass{aa}

\usepackage{graphicx}
\usepackage{txfonts}
\usepackage{natbib}
\bibpunct{(}{)}{;}{a}{}{,} 
\usepackage{hyperref}
\hypersetup{
    colorlinks=true,
    linkcolor=blue,
    citecolor=blue,
    }

\begin{document} 


\title{Influence of the turbulent magnetic pressure on isothermal jet emitting disks}


\author{N. Zimniak\inst{1}, J. Ferreira\inst{1} \and J. Jacquemin-Ide\inst{2}}

\institute{Univ. Grenoble Alpes, CNRS, IPAG, 38000 Grenoble, France\\
    \email{nathan.zimniak@univ-grenoble-alpes.fr}
    \and 
    Center for Interdisciplinary Exploration \& Research in Astrophysics (CIERA), Evanston, IL 60202, USA
}

\date{Received 25 April 2024 / Accepted 27 October 2024}

\abstract
{The theory of jet emitting disks (JEDs) provides a mathematical framework for a self-consistent treatment of steady-state accretion and ejection. A large-scale vertical magnetic field threads the accretion disk where magnetic turbulence occurs in a strongly magnetized plasma. A fraction of mass leaves the disk and feeds the two laminar super-Alfv\'enic jets. In previous treatments of JEDs, the disk turbulence has been considered to provide only anomalous transport coefficients, namely magnetic diffusivities and viscosity. However, 3D numerical experiments show that turbulent magnetic pressure also sets in.}
{We analyze how this turbulent magnetic pressure modifies the classical picture of JEDs and their parameter space.} 
{We included this additional pressure term using a prescription that is consistent with the latest 3D global (and local) simulations. We then solved the complete system of self-similar magnetohydrodynamic (MHD) equations, accounting for all dynamical terms. The magnetic surfaces are assumed to be isothermal, limiting the validity of our results to cold outflows. We explored the effects of the disk thickness and the level of magnetic diffusivities on the JED response to turbulent magnetic pressure.}
{The disk becomes puffier and less electrically conductive, causing radial and toroidal electric currents to flow at the disk surface. Field lines within the disk become straighter, with their bending and shearing occurring mainly at the surface. Accretion remains supersonic, but becomes faster at the disk surface. Large values of both turbulent pressure and magnetic diffusivities allow powerful jets to be driven, and their combined effects have a constructive influence. Nevertheless, cold outflows do not seem to be able to reproduce mass-loss rates as large as those observed in numerical simulations.}
{Our results are a major upgrade of the JED theory, allowing a direct comparison with full 3D global numerical simulations. We argue that JEDs provide a state-of-the-art mathematical description of the disk configurations observed in numerical simulations, commonly referred to as magnetically arrested disks (MADs). However, further efforts from both theoretical and numerical perspectives are needed to firmly establish this point.}

\keywords{
        Accretion, accretion disks --
        Magnetohydrodynamics (MHD) --
        Turbulence --
        ISM: jets and outflows
}

\titlerunning{Turbulent magnetic pressure in isothermal JEDs}
\authorrunning{N. Zimniak et al.}
\maketitle


\section{Introduction}
Accretion disks are found in a wide variety of astrophysical objects: around supermassive black holes (BHs) at the center of active galaxies (AGNs) and quasars \citep{blan19,lu23}, around every low-mass young stellar object (YSO) where planet formation occurs \citep{agra14,deva20}, around interacting binary systems hosting either a white dwarf (cataclysmic variables), a neutron star or a BH (X-ray binaries) \citep{ponti16,tudor17}, or even some post-AGB stars \citep{gorl15,boll17}. For many years, accretion was believed to be mainly driven by turbulence, which would allow an outward radial transport of the disk angular momentum, while mass would accrete onto the central object \citep{shak73}. The quest for the instability that would allow nearly Keplerian accretion flows to become turbulent led to the conclusion that accretion disks must be magnetized \citep{balb91}. The magneto-rotational instability (MRI) is indeed a very robust and fast (growing on dynamical timescales) instability that leads to fully developed 3D turbulence \citep{balb03}. It has thus become clear that accretion disks must be described as magnetized plasmas, namely within the framework of magnetohydrodynamics (MHD).

Simultaneously, highly collimated bipolar supersonic jets have also been discovered emerging from the same astrophysical objects. Mostly detected in the radio band around compact objects or in radio to optical emission lines in YSOs, they share the property of being emitted from the innermost regions close to the central object. Moreover, both compact sources and YSOs show a tight correlation between accretion and ejection signatures \citep{corb13,nisi18}. These properties appear to be universal, as they do not depend on the nature of the central object  (BH or stellar object), thus favoring the scenario where jets are emitted from the accretion disks. Since, in all resolved sources, jet collimation appears to be effective already very close to the source, jet acceleration and collimation must work hand in hand. This can occur naturally if a large-scale vertical magnetic field defines a magnetic channel along which the ejected plasma flows. Such a solution was first proposed by \citet{love76} and \citet{blan76}, and the coupled acceleration and collimation dynamics of jets emitted from the surface of a Keplerian accretion disk was later demonstrated by \citet{blan82}.

There have been many attempts to analytically link accretion disks to magnetized outflows, and it is beyond the scope of this introduction to cite them all. Most models either rely on strong approximations (e.g., neglected forces, negligible wind mass loss, and simplified wind equations, \citealt{ward93, nara95a,li95,ogil98,bai16a} to cite only a few) or use a self-similar variable separation approach that allows all dynamical terms to be included in the equations \citep{ferr93a,ferr95}. 

Only the anomalous turbulent coefficients need to be specified to close the system of equations describing the accretion-ejection theory. The turbulent transport coefficients resulting from MRI turbulence can be derived from measurements conducted in 3D shearing box simulations \citep{hawl95,salv16} or 3D global simulations \citep{zhu18,jacq21}. The \cite{shak73} angular momentum transport coefficient has indeed been computed numerous times, and it is a well-constrained function of the local magnetic field strength \citep{salv16}. In contrast, turbulence resistivity has been measured sparingly, and no clear dependency on the field strength has been identified \citep{lesu09,guan09,from09}. The situation becomes even more obtuse when pertaining to the complete resistivity tensor \citep{gres15}.

Magnetized accretion-ejection structures describe both the (turbulent) viscous and resistive MHD disk and its two (laminar) ideal MHD jets as a single, mathematically connected, system. The smooth crossing of the Alfv\'en point determines the position of this critical point as well as the strength of the toroidal field at the disk surface. This, in turn, fixes not only the torque allowing accretion but also the vertical magnetic compression acting on the disk, and hence the disk magnetization $\mu$ that is consistent with the disk ejection efficiency $\xi$ and the smooth crossing of the slow-magnetosonic critical point (see \citealt{ferr97} for details).

The goal of the theory is then to provide the relation $\xi(\mu)$, where the disk ejection efficiency is defined as $\dot M_a (r) \propto r^\xi$ and $\dot M_a (r)$ is the disk accretion rate measured at a given radius. 
The theory aims to directly link the properties of the disk to the outflow. This is crucial for understanding the long-term evolution of an accretion disk and connecting observed outflow properties to the disk itself. Models of disk evolution, considering mass loss through outflows, must directly prescribe the ejection efficiency, which would otherwise remain unconstrained without an accretion-ejection theory \citep{tabone_secular_2022}. Ejection efficiency is also vital for comparing theories of magnetized outflows to observations of disk winds in low-mass X-ray binaries (LMXRBs), leading to constraints on disk properties \citep{chakravorty_magneto_2016,chakravorty_absorption_2023,ranjan_datta_impact_2024}. Moreover, the mass ejection index significantly influences the history of X-ray binaries (XRBs) by altering mass and angular momentum loss, thus affecting the secular evolution of orbital separation \citep{gallegos-garcia_angular_2023}.


The first  self-consistent accretion-ejection solutions were found for isothermal (\citealt{ferr97}, hereafter F97) or adiabatic \citep{cass00a} cold outflows. The disk magnetization $\mu$ was found to lie in a very narrow range, between $0.1$ and $0.8$, with a typical disk ejection efficiency $\xi \sim 0.01$, leading to very fast and tenuous collimated jets. The accretion disk associated with these solutions has thus been termed jet emitting disk \citep[JED,][]{ferr06a}. In a JED, accretion is supersonic and the disk is much less radiative than any other usual accretion disk solution (due to energy transfer to jets), with important implications for instance in X-ray binary cycles (\citealt{ferr06a, marc18a,marc22} and references therein). We note that the presence of supersonic accretion was subsequently found in 3D global simulations of highly magnetized disks \citep{jacq21,scep24}.


Until recently, the minimum value of the disk magnetization was constrained to order equipartition within the JED theory.
At $\mu < 10^{-2}$, MRI sets in despite the presence of anomalous diffusivities, and solutions display spatial (vertical) oscillations. Such oscillations are actually nonlinear channel modes due to the MRI \citep{jacq19}. Despite a smaller magnetic field, these oscillations help to drive the outflow by building up a stronger toroidal field at the disk surface. Since the vertical pinching of the disk is smaller, these solutions allow for denser isothermal (cold) outflows with a typical ejection efficiency $\xi \sim 0.1$. In this case, the denser outflow is also much slower, so the accretion disk associated with these weakly magnetized solutions has been termed wind emitting disk (WED). In a WED, accretion is always subsonic and the magnetic configuration is reminiscent of the so-called magnetic towers \citep{lynd03}. As discussed in \cite{jacq19}, it is doubtful that oscillations would survive within the disk, since secondary instabilities (such as Kelvin-Helmholtz or Rayleigh-Taylor) would most likely be triggered, leading to enhanced anomalous transport within the turbulent region. This is therefore a good incentive to look at global 3D numerical solutions where MHD turbulence is fully resolved. Another reason to look at global simulations is that local shearing box simulations are subject to several biases when simulating stratified disks with vertical magnetic fields, which could skew the vertical profiles of the anomalous transport coefficients \citep{lesur_magnetorotational_2013}.

Nowadays, general relativistic magnetohydrodynamic (GRMHD) simulations are commonly categorized using the magnetically arrested disk (MAD) and the standard and normal evolution (SANE) terminology (e.g., \citealt{akiy21}). A MAD is defined by the magnetic flux at the horizon of the BH\footnote{This terminology goes back to the work of \citet{nara03} and \citet{igum03}, who argued that if the magnetic field was too large, the disk would not rotate anymore and would therefore be arrested. But it turned out that MADs are actually rotating, turbulent, and launching MHD outflows; facts that put into question the chosen terminology \citep{McKi12,bege22}.}. Specifically, it is considered that the system enters the MAD state when the flux normalized by the accretion rate reaches its maximal value (see e.g., \citealt{tchekhovskoy_efficient_2011,nara22} and references therein).
On the contrary, whenever the magnetic flux on the BH is less than that specific value, the numerical outcome has been termed SANE. 
However, the numerical convergence to one state or another depends mostly on two things: (1) the initial conditions (how much initial magnetic flux is available in the computational domain) and (2) how long the simulation runs (since the flux appears to be advected inward, leading to a growth of the magnetization of the inner regions). The reason why the definitions of the MAD and SANE state rely on the dynamics of the inner BH and not on the disk structure is twofold: 1) for most BH spin, the \cite{blan77} jet energetically outshines any contribution from the disk; and 2) due to numerical limitations, only the inner regions, $r<20\,\,r_g$, would have achieved a complete steady state, making a study of the large-scale turbulent disk dynamics out of reach. There is no explanation yet of the physical state reached by the accretion disk in each case. It is nevertheless tempting to associate numerical MADs and SANEs with their semi-analytical counterparts, JEDs and WEDs, due to them being associated with different field strengths.

It is only recently that the turbulent structure of a  disk threaded by a large-scale vertical field has started to be thoroughly analyzed \citep{scep24,mani24}, following similar efforts in nonrelativistic simulations \citep{zhu18,jacq21}. Those global numerical investigations on the turbulent structure have shown that the role played by the turbulent magnetic pressure is of paramount importance in shaping the disk vertical balance \citep{jacq21,scep24}. This is a major effect that has not yet been studied in semi-analytical approaches of magnetized accretion-ejection structures. 


The aim of this paper is precisely to investigate the influence of turbulent magnetic pressure in highly magnetized accretion disks, namely in JEDs. Section~2 recalls the assumptions, equations, and parameters needed to fully describe JEDs. In particular, we discuss how MHD turbulence is taken into account in our self-similar steady-state mean-field approach. Section~3 focuses on analyzing the impact of this additional pressure term on the disk structure, by considering only the slow magnetosonic constraint. The full parameter space, including the additional Alfv\'enic constraint, is analyzed in Section~4. While Sections~3 and 4 focus on a fiducial set of parameters, in Section~5 we vary the disk thickness and the level of the magnetic diffusivities in order to explore their effects on the JED response to a turbulent magnetic pressure. Section~6 summarizes our results and discusses their implications, in particular providing comparisons with some published results of 3D numerical simulations. Finally, we conclude in Section~7.


\section{Accretion-ejection theory}

\subsection{Magnetohydrodynamic equations} 

The accretion-ejection theory assumes axisymmetry and stationarity. In this framework, a disk of plasma orbits in near-Keplerian motion around a central object of mass $M$. The disk self-gravity is neglected and the gravitational potential of the central object is Newtonian. The disk is assumed to be threaded by a large-scale vertical magnetic field. The whole system is assumed to have a plane symmetry so that $z=0$ describes the disk midplane and the bipolar jets are symmetric. Using the cylindrical coordinates $(r, \phi, z)$ the plasma velocity and magnetic field are decomposed into poloidal and toroidal components, namely $\Vec{u} = \Vec{u}_p + \Omega r \Vec{e}_{\phi}$ and $\Vec{B} = \Vec{B}_p + B_{\phi} \Vec{e}_{\phi}$, where $\Omega$ is the plasma angular velocity. The poloidal magnetic field writes
\begin{equation}
\Vec{B}_p = \frac{\nabla a}{r} \times \Vec{e}_\phi \, ,
\end{equation}
\noindent where $a(r,z)$ is the magnetic flux function ($a(r,z)=a_o$ describes a magnetic surface of constant poloidal magnetic flux). Hence, a jet can be seen as magnetic surfaces anchored over a radial extent of the accretion disk and nested around each other.  

The set of partial differential equations describing a steady-state magnetized accretion-ejection structure writes 
\begin{align}
   &\Vec{\nabla} \cdot \left( \rho \Vec{u}_p \right) = 0 \label{massconservation} \\[0.5em]
   &\rho \left( \Vec{u} \cdot \Vec{\nabla} \right) \Vec{u} = - \Vec{\nabla} \left( P + P_{turb} \right) + \rho \Vec{\nabla} \Phi_G + \Vec{J} \times \Vec{B} + \Vec{\nabla} \cdot \Vec{\mathcal{T}} \label{momentumconservation} \\[0.5em]
   &\eta_m J_{\phi} \Vec{e}_{\phi} = \Vec{u}_p \times \Vec{B}_p \label{ohmlaw} \\[0.5em]
   &\Vec{\nabla} \cdot \left( \frac{\nu_m^{\prime}}{r^2} \Vec{\nabla} \left[ r B_{\phi} \right] \right) = \Vec{\nabla} \cdot \left( \frac{1}{r} \left[ B_{\phi} \Vec{u}_p - \Omega r \Vec{B}_p \right] \right) \label{inductioneq} \\[0.5em]
   &P = \rho C_s^2 \, ,    \label{eos}
\end{align}
\noindent and accounts for, respectively, mass and momentum conservation, Ohm's law, toroidal magnetic field induction and the equation of state \citep{ferr95,cass00a}. In these equations, $\rho$ is the plasma density, $P$ the total (gas + radiation)  pressure,  $\Phi_G = - GM/\sqrt{r^2 + z^2}$ the gravitational potential of the central object, $\Vec{J} = \Vec{\nabla} \times \Vec{B}/\mu_0$ the electric current density and $C_s$ the sound speed. In this paper, radiation pressure is neglected and the plasma is assumed to be thermalized (identical electronic and ionic temperatures) so that $C_s^2= 2 \rho kT/m_p$, where $m_p$ is the proton mass. An energy equation should then be provided to compute the plasma temperature $T$. However, we will only focus on isothermal structures where $T$ remains constant along magnetic field lines but is allowed to vary radially.

The disk is expected to be fully turbulent \citep{balb03}, so the above equations must be understood as a mean-field approach only. Such turbulence is further assumed to give rise to enhanced (anomalous) transport coefficients such as a viscosity $\nu_v$, magnetic diffusivity in the poloidal plane $\nu_m$ (and resistivity $\eta_m= \mu_o \nu_m$, where $ \mu_o $ is the vacuum permeability) and in the toroidal direction $\nu'_m$. The presence of an anomalous viscosity implies the existence of a viscous stress tensor $\Vec{\mathcal{T}}$ (see next section for its expression). The novelty of this study is the inclusion of a turbulent magnetic pressure $P_{turb}$ in Eq.~\eqref{momentumconservation}. As we will show, this term indeed plays a major role and brings about significant changes in the overall accretion-ejection picture.

Magnetohydrodynamic turbulence is expected to decline vertically as the ejected plasma moves away from the disk. By definition, the flow transits from resistive MHD to ideal MHD at the disk surface, allowing for a smooth connexion to a super-magnetosonic (hereafter super-SM) and, further out, super-Alfv\'enic (hereafter super-A) outflow. Once in ideal MHD, the above equations lead to (F97)
\begin{align}
&\Vec{u}_p = \frac{\eta(a)}{\mu_o \rho}\Vec{B}_p  \label{eta} \\
&\Omega_*(a) = \Omega - \eta \frac{B_{\phi}}{\mu_0 \rho r} = \frac{\Omega}{(1 - g)},\, \,  \mbox{  with   } g = \frac{m^2}{m^2-1}\left ( 1 - \frac{r_A^2}{r^2}\right ) \label{Omegastar} \\
&L(a) = \Omega_* r_A^2 = \Omega r^2 - \frac{r B_{\phi}}{\eta} \label{L} \\
&E = \frac{u^2}{2} + H + \Phi_G - \Omega_* \frac{r B_{\phi}}{\eta} + E_{turb} \, , \label{E}
\end{align}
\noindent where  $\eta(a)$ is the mass flux to magnetic flux ratio, $\Omega_*(a)$ the rotation rate of the magnetic surface, $L(a)$ the total specific angular momentum carried along that surface, $m = u_p/V_{Ap}$ is the poloidal Alfv\'enic Mach number, $r_A$ the cylindrical radius where the poloidal flow speed meets the poloidal Alfv\'en speed (A critical point, $m=1$). These 3 quantities are MHD invariants along any magnetic surface anchored on a radius $r_o$ at the disk midplane of constant magnetic flux $a$ (along with the plasma temperature in our isothermal case).

The last quantity is the total specific energy $E$ and is also an MHD invariant (Bernoulli integral) along the flow whenever energy is neither gained nor lost (see e.g., \citealt{cass00b} when additional heating is added). It states that jet acceleration is a conversion of the (initially dominant) magnetic energy and enthalpy reservoir $H$ into jet kinetic energy. In this paper, since turbulent pressure has been added, there is an extra energy reservoir 
\begin{equation}
E_{turb}= -  \int_{s^+}^s \frac{\partial P_{turb}}{\partial s}\frac{ds}{\rho} \, ,
\end{equation} 
\noindent where the integration starts from the disk surface (taken as the transition $s^+$ between the resistive and the ideal MHD regime) and is done with the curvilinear coordinate $s$ along a given magnetic surface. Since the turbulence is assumed to decrease vertically, $E_{turb}>0$. However, it turns out to be always negligible in all the cases studied here so that it can be safely discarded, and using $E(a)$ as an invariant is not too bad an approximation.

\begin{figure*}
    \centering
    \resizebox{0.8\hsize}{!}{\includegraphics{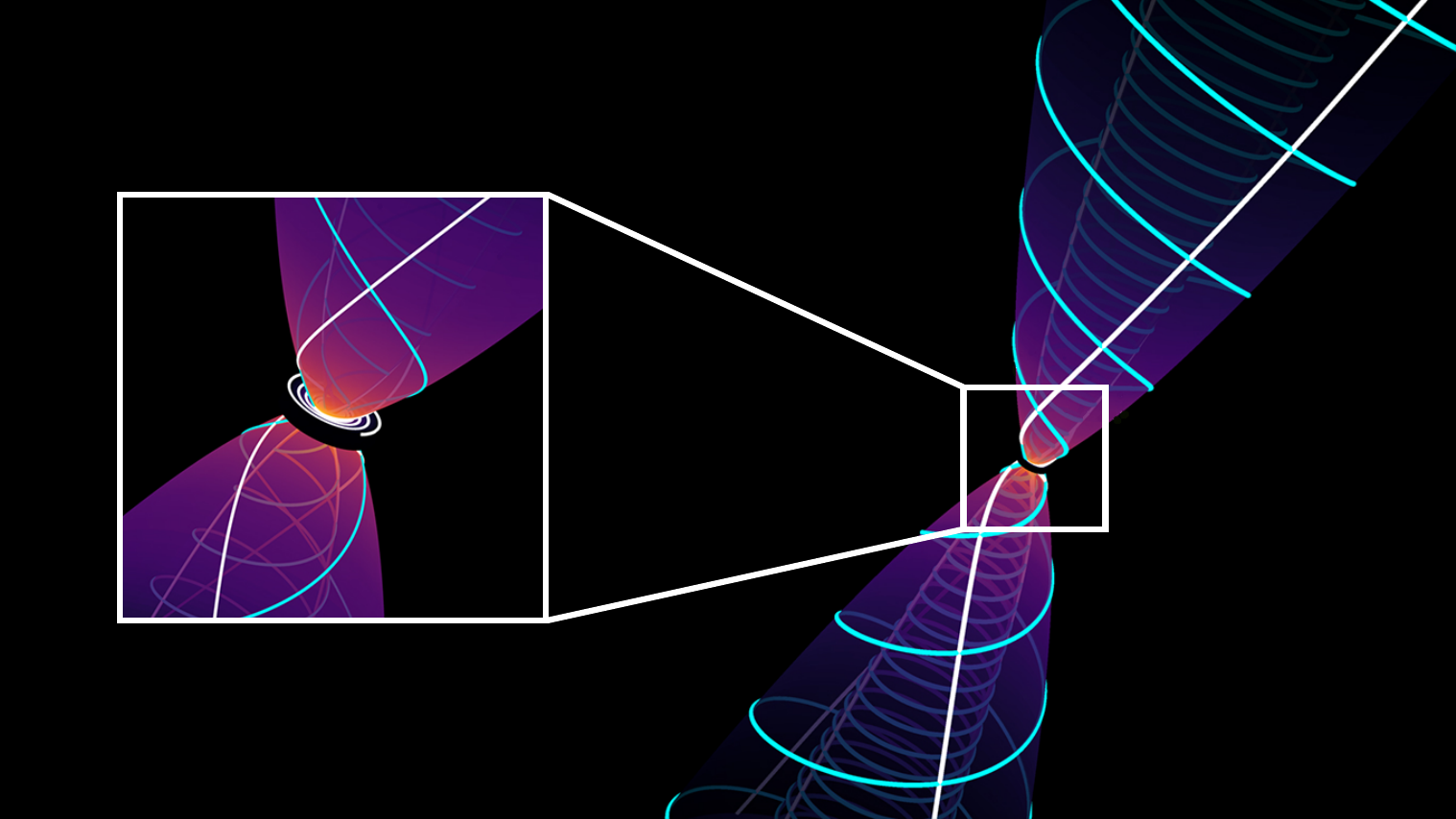}}
    \caption{3D view of a typical super-A JED solution. The blue lines show the magnetic field lines and the white lines show the streamlines along the same magnetic surfaces, anchored at $r_0 = 1.0$ and $r_0 = 5.0$.}
    \label{fig:jed_3d}
\end{figure*}

\subsection{MHD turbulence prescriptions} 

In the standard accretion disk theory, turbulence leads to an anomalous viscosity $\nu_v= \alpha_v C_s h$, where $\alpha_v$ is a free parameter of the model \citep{shak73}. Building upon this idea and anticipating that turbulence should be of MHD origin, \citet{ferr93a} proposed that the anomalous poloidal magnetic diffusivity in the disk midplane should be written as
\begin{equation}
    \nu_m = \alpha_m V_A h \, .\end{equation}
\noindent Such a scaling allows to link the two transport coefficients by introducing the (effective) magnetic Prandtl number 
\begin{equation}
{\cal P}_m = \frac{\nu_v}{\nu_m} = \frac{\alpha_v}{\alpha_m} \mu^{-1/2} \, .
\end{equation}
It turns out that MRI-driven MHD turbulence 1) can indeed be safely described by a viscous stress tensor $\Vec{\mathcal{T}}$ with a dominant component $\mathcal{T}_{r\phi} = \rho \nu_v r \frac{\partial \Omega}{\partial r}$ (\citealt{balb03} and references therein) and 2) that the Shakura-Sunyaev parameter follows the scaling
\begin{equation}
\alpha_v \simeq 8 \mu^{1/2} \, ,
\label{eq:salv}
\end{equation}
\noindent at least in shearing box simulations \citep{hawl95, salv16}. The prescriptions used in the theory are therefore consistent with MRI-driven turbulence and lead to $\alpha_m {\cal P}_m \sim 8 $, namely a true constant independent of the field strength. Note however that its value seems to depend on simulation details such as disk aspect ratio or local (shearing box) versus global simulations. We will therefore assume that a product $\alpha_m {\cal P}_m$ around unity is acceptable. 

Measuring the magnetic diffusivity in 3D simulations is a very difficult task and has been only rarely done. But both local \citep{lesu09,guan09,from09} and global \citep{zhu18,jacqT21,jacq21} simulations tend to show that ${\cal P}_m \sim 1-3$. For the rest of the paper, and for the sake of simplicity, we will use ${\cal P}_m = 1$ and allow $\alpha_m$ to vary (F97, \citealt{jacq19}).

However, it is still unclear whether MHD turbulence in accretion disks leads to a fully isotropic resistivity tensor, that is, if the turbulent diffusion is the same in all directions (see however \citealt{lesu09,gres15}). \citet{ferr95} showed that some slight anisotropy is needed in order to provide the best conditions for cold jets to be launched. This translates into a diffusivity $\nu'_m$ in the toroidal direction possibly larger than $\nu_m$, with the introduction of an anisotropy parameter
\begin{equation}  
\chi_m = \frac{\nu_m}{\nu'_m}
\end{equation}
\noindent (see e.g., the parametric studies in \citealt{cass00a, jacq19}). Nevertheless, and again for the sake of simplicity, we will keep $\chi_m = 1$ in our study.

The existence of a turbulent magnetic pressure in disks is a natural outcome of MRI. Its magnitude (as measured at the disk midplane) depends of course on the initial (vertical) magnetic field. Following \citet{salv16}, who provided analytical fits from local simulations, we will use the scaling
\begin{equation}
    P_{turb} = \frac{<\delta B^2>}{2\mu_o} =\rho \delta V_A^2 = \alpha_P \sqrt{\mu} P \, , \label{eqPturb}
\end{equation}
\noindent at the disk midplane, where $<\delta B^2>$ is some local average. The fits made in \citet{salv16} provided a value $\alpha_P \sim 26$ which is quite high and shows how important this component is in local (shearing box) simulations. This scaling has been validated in both global MHD and GRMHD simulations at low magnetization values \citep{jacq21,mishra20}, and has been observed in a global MHD simulation at higher magnetization values \citep{jacqT21}. The values of $\alpha_P$ derived from these simulations usually tend to be smaller and may also depend on $h/r$ (e.g., $\alpha_P \sim 4$; \citealt{jacqT21}). We will therefore use $\alpha_P$ as a free parameter. Note that in our axisymmetric approach, this turbulent pressure only plays a role in the radial and the vertical momentum equations. In agreement with numerical simulations, we expect the dominant contribution to this pressure to arise from the toroidal magnetic field. This allows therefore to use the same term in both equations.

Since this turbulent pressure adds up with the kinetic pressure, we expect an easier vertical balance against gravity and laminar magnetic compression or, alternatively, some enhanced mass-loss rate from the disk. In the radial direction, the turbulent pressure should also provide some support against gravity. Both situations actually correspond to a modification of the local sound speed. Indeed, since the prescription amounts to $ P_{turb} \propto P \propto \rho$, the modified sound speed writes
\begin{equation}
\tilde C_s^2 =\left ( 1 + \alpha_P \mu^{1/2} \right )  C_s^2 = C_s^2 + \delta V_A^2 \, .
\end{equation}
This modified sound speed can actually be interpreted as some fast magnetosonic speed. While in weakly magnetized WEDs compressible turbulence does not cause a significant deviation of the sound speed, this is no longer the case in strongly magnetized JEDs. As a consequence, taking this turbulence into account shifts to higher altitudes the (modified) sonic critical point that is encountered in resistive MHD \citep{ferr95}.

As the matter leaves the disk and enters the ideal MHD regime, the turbulence is assumed to fade away, leading to negligible anomalous transport coefficients. In practice, the resistive-ideal MHD transition occurs only when the LHS resistive terms in Eqs.~(\ref{ohmlaw},\ref{inductioneq}) as well as the viscous torque in Eq.~(\ref{momentumconservation}) become smaller than 1\% of the others. But there is no physical reason to neglect the turbulent magnetic pressure, at least at the base of the ideal region. We therefore keep the turbulent term in the ideal MHD equations and this is the reason why the Bernoulli integral (Eq.~\ref{E}) has a turbulent contribution. This also introduces a modification of the SM and FM critical speeds defined in ideal MHD regime, which can be written as
\begin{align}
   &V_{SM, FM}^2 = \frac{1}{2} \left( \tilde C_s^2 + V_{A}^2 \pm \sqrt{\left(  \tilde C_s^2 + V_{A}^2 \right)^2 - 4  \tilde C_s^2 V_{An}^2 } \right) \nonumber \\
   &V_{A}^2 = V_{Ap}^2 + V_{A\phi}^2 \label{eqvsm} \\
   & V_{An}^2 = \frac{ \left( \Vec{B}_p \cdot \Vec{n} \right)^2 }{\mu_0 \rho} \, , \nonumber 
\end{align}
\noindent where the SM (resp. FM) speed is defined with the minus (resp. plus) sign, $V_A$ is the total Alfv\'en speed and $V_{An}$ is the projection of the Alfv\'en speed in the self-similar direction $\Vec{n} $ (see \citealt{ferr95} and references therein for more details). Thus, the turbulent magnetic pressure could potentially play an important role at the base of the jet where the flow must become super-SM if it introduces a significant modification of the critical phase speed $V_{S\!M}$. In any case, due to a prescribed decaying vertical profile of the turbulence (see below), such an effect remains quite negligible in the study done in this paper.

\subsection{Self-similar ansatz} 


\begin{figure}
      \centering
      \resizebox{0.8\hsize}{!}{\includegraphics{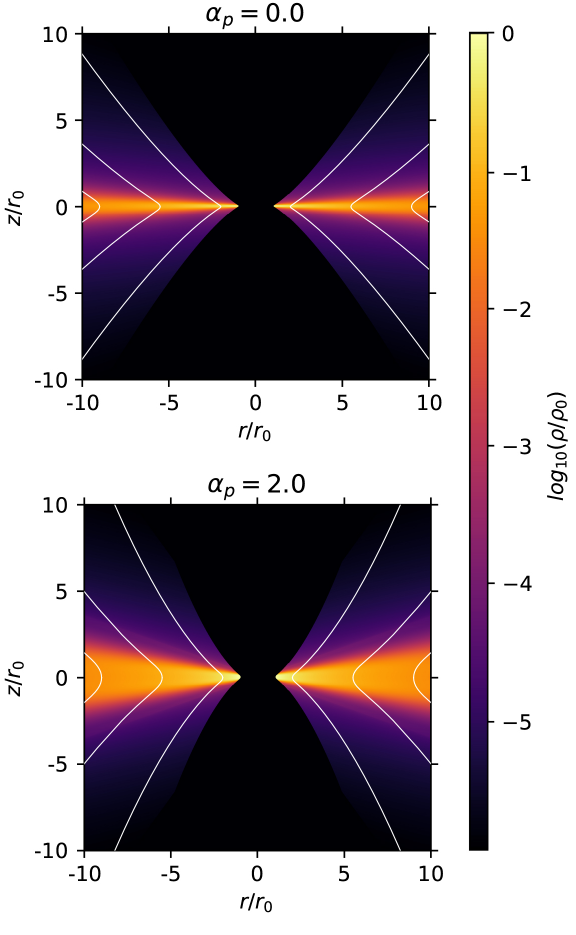}}
      \caption{Poloidal cross-section of two super-SM JED solutions with the same $\xi = 3 \times 10^{-3}$, obtained with $\epsilon = 0.1, \alpha_m =\mathcal{P}_m =\chi_m= 1$, without ($\alpha_P=0$, top panel) and with ($\alpha_P=2$, bottom panel) a turbulent magnetic pressure. The background color represents the logarithm of the density normalized to its midplane value and the white lines show the magnetic surfaces. See text for more details.}
      \label{fig:2D_disk}
\end{figure}


The set of MHD equations \eqref{massconservation}-\eqref{eos} forms a system of nonlinear partial differential equations. Solving such a system while keeping all dynamical terms in the equations can be done by a variable separation method \citep{ferr93a,ferr95}. Since gravity is indeed the dominant contribution, a steady-state solution over a large radial extension can only be obtained if each term follows the same scaling with the cylindrical variables $(r,z)$. Therefore, each physical quantity $A(r,z)$ is assumed to follow the self-similar form
\begin{equation}
    A(r,z) = A_o \left( \frac{r}{r_o} \right)^{\zeta_A} f_A (x) \, ,
\end{equation}
\noindent namely a power law of the radius with exponent $\zeta_A$ times a function $f_A (x)$ of the self-similar variable $x=z/h(r)$, with a disk aspect ratio $\epsilon= h/r$ a constant of the radius. Actually, keeping all the dynamical terms in the equations leads to the mathematical constraint that {\it all} dimensionless quantities must remain constants of the radius. But this is the price to pay for obtaining exact MHD solutions. The subscript "$o$" stands for a quantity evaluated at the disk mid-plane $x = 0$. For instance, the magnetic flux function writes $a(r,z) = a_o (r/r_o)^\beta \Psi(x)$ so the shape of a magnetic surface with constant magnetic flux $a(r,z)=a_o$ is given by $ r= r_o \Psi^{-1/\beta}$, where $r_o$ is the anchoring radius at the disk midplane.

The self-similar ansatz replaces the system \eqref{massconservation}-\eqref{eos} with a system of algebraic equations on the radial exponents $\zeta_A$ and a system of nonlinear ordinary differential equations (ODEs) on the functions $f_A(x)$. All the exponents $\zeta_A$ are then expressed as a function of the disk ejection index $\xi$, namely
\begin{equation}
\begin{array}{lll}
\dot{M}_a \propto r^{\xi}  &  a \propto r^{\frac{3}{4} + \frac{\xi}{2}} & u_i \propto r^{-1/2}  \\
\rho  \propto r^{\xi - 3/2}    &  B_i \propto r^{\frac{-5}{4} + \frac{\xi}{2}}    & T  \propto r^{-1} \, ,
\end{array}
\label{eq:exposants}
\end{equation}
\noindent where the subscript "$i$" stands for the three components ($r,\phi,z$). These exponents are important because they tell us that the magnetic field radial distribution must be consistent with the density distribution, which of course must be closely related to the disk mass-loss rate. Any discrepancy between these exponents unavoidably leads to a time-dependent evolution of the system. This has to be kept in mind when comparing them to the outcome of numerical simulations for instance.

According to the self-similar approach, vertical profiles must also be assumed for all turbulent quantities. This is probably the least controlled prescription, as they have been very seldom provided in the literature. Moreover, the local MHD turbulence and its associated vertical profiles are most certainly self-adapting to the global accretion-ejection interrelations, while in our approach they must be prescribed ab initio without any feedback from the laminar solution. Deeper investigations using vertical profiles derived from global 3D simulations will be done in future work. Here, we limit ourselves to a simple Gaussian profile $m(x)= e^{- x^2/2x_t^2}$ of scale $x_ t$ as done previously \citep{ferr93a,ferr95,cass00a, jacq19}. This leads to a decrease of all turbulence-related terms on a scale $x_t \sim 1$. More precisely, the self-similar profiles of the magnetic diffusivities, viscous torque and turbulent magnetic pressure follow  
\begin{align}
   &\nu_m \propto \nu'_m \propto m(x) \nonumber \\
&\frac{1}{r^2} \frac{\partial}{\partial r} \left ( r^2 \mathcal{T}_{r\phi} \right) \propto m(x) f_\rho(x)   \\
   & P_{turb} \propto m(x)f_P(x) \, . \nonumber
\end{align}

\noindent However, such simplified profiles are roughly consistent with simulations of strongly magnetized accretion disks, namely JEDs with $\mu \sim 1$ \citep{jacq21,scep24}. On the contrary, the vertical profiles seen in weakly magnetized accretion disks, WEDs with $\mu \sim 10^{-3}$ or less, are clearly different from simple Gaussians \citep{zhu18,jacqT21}. This is consistent with the finding that decaying Gaussian profiles lead to vertical oscillations of all quantities in WEDs, as discussed in \citet{jacq19}. For this reason, this paper will focus only on JEDs, that is, solutions without vertical oscillations.

\subsection{Summary of parameters and methodology} 

We numerically solve the system of ODEs on the functions $f_A$ using a predictor-corrector integrator for stiff equations. Starting from the disk midplane, the integration propagates upward using first the resistive MHD equations. When the conditions for ideal MHD are satisfied (resistive and viscous torque negligible wrt the other terms), we switch to the ideal MHD equations. No significant jumps are seen in this transition. Once in ideal MHD, the plasma poloidal velocity is parallel to the poloidal magnetic field and the flow must cross the SM critical point. The regularity condition that allows a smooth transition to the super-SM regime fixes a quantity defined at the disk midplane. We choose to adjust the disk magnetization $\mu$ since it finely controls the disk vertical balance for a given ejection efficiency $\xi$ (see \citealt{ferr95} for more details).
 
As the super-SM flow is further accelerated (mostly by magnetic means, since our isothermal outflows have a negligible enthalpy), it must become super-A. Again, this is done by fine-tuning another physical quantity defined at the disk midplane. We choose to adjust the curvature of the poloidal field lines (parameter $p$ associated with the toroidal electric current density $J_{\phi o}$), which is loosely related to the magnetic field bending at the disk surface. Once a new value of $p$ has been chosen, the whole procedure of finding the new $\mu$ allowing for trans-SM solutions must be repeated (see F97 for more details). An example of a typical JED super-Alfvénic solution is illustrated in Fig~\ref{fig:jed_3d}.

We are not looking for super-FM outflows. Not only is this last critical point deeply biased by the self-similar ansatz (unlike the first two, see discussions in \citealt{blan82, ferr95, bogo97, vlah00}), but the regularity condition can be satisfied by adjusting the jet adiabatic index beyond the Alfv\'en point and thus does not provide a strong constraint on the disk physics (see \citealt{ferr04}). In any case, all super-A solutions become super-FM in the conventional sense, that is, the poloidal kinetic energy always becomes larger than the total magnetic energy (F97).
 
In summary, a magnetized accretion-ejection solution leading to super-A isothermal ($\gamma=1$) outflows is described by the set of six parameters
\begin{equation}
    \begin{aligned}
        &\xi = \frac{dln \left( \dot{M}_a \right)}{dln \left( r \right)}
        & \qquad
        &\epsilon = \frac{h}{r}
        & \qquad
        &\alpha_m = \frac{\nu_m}{V_A h}
        \\
        &\chi_m = \frac{\nu_m}{\nu_m^{\prime}}
        & \qquad
        &\mathcal{P}_m = \frac{\nu_v}{\nu_m}
        & \qquad
        &\alpha_P = \frac{P_{turb}}{\sqrt{\mu} P}
    \end{aligned}
\end{equation}
\noindent defined at the disk midplane. Four of them ($\alpha_m, {\cal P}_m, \chi_m, \alpha_P$) describe the MHD turbulence and are therefore unavoidable, unless future work on MHD turbulence provides their scaling as a function of, for instance, the disk aspect ratio $\epsilon$ and the disk magnetization $\mu$. The disk aspect ratio $\epsilon$ is a real free parameter since no proper energy equation is solved. This is a drawback of the theory, but it guarantees the use of self-similarity and allows us to gain valuable insight into the effect of disk thickness on jet launching.

The physical ingredient that allows both accretion and super-A ejection is the vertical {\it laminar} magnetic field, whose strength is measured at the disk midplane by the disk magnetization parameter 
\begin{equation}
\mu = \frac{V_A^2}{C_s^2} = \frac{B_{zo}^2}{\mu_o P_o} \, .
\end{equation}
\noindent The theory thus provides the disk ejection efficiency $\xi(\mu)$ for a set of prescribed values ($\epsilon, \alpha_m, {\cal P}_m, \chi_m, \alpha_P$). In practice, all possible values for $\xi$ between $10^{-4}$ and $1$ are scanned (F97), looking for the value $\mu$ (and curvature $p$) that leads to super-A outflows.  
 
The aim of this work is to study the effect of turbulent magnetic pressure on isothermal JED solutions. This is done in two steps. First (Sect.~\ref{sec:super-SM}), we look at how both the JED vertical and radial balance respond to an increasingly large $\alpha_P$. Since strict stationarity requires satisfying two constraints (SM and A), it is important to assess how each of them responds to the presence of turbulent pressure in the disk. Moreover, the crossing of the A point may be strongly affected by heat deposition or pressure effects above the disk \citep{ferr04}, both effects being neglected in our isothermal approach. In a second step (Sect.~\ref{sec:super-A}) we consider both SM and A constraints, thus providing the final parameter space and its modification with $\alpha_P$.

The analyzes carried out in Sections~\ref{sec:super-SM} and \ref{sec:super-A} are performed for our fiducial set ($\epsilon=0.1, \alpha_m = {\cal P}_m = \chi_m=1$), while $\alpha_P$ is freely varied. Section~\ref{sec:param} explores the effect of $\alpha_P$ for different disk thicknesses $\epsilon=0.01, 0.1, 0.2$ and for different turbulence levels $\alpha_m=0.5, 1, 3$.

\section{Impact on the disk structure: super-SM parameter space}
\label{sec:super-SM}

In this section, we investigate the effect of the turbulent magnetic pressure on the JED internal structure for our fiducial parameter set defined by ($\epsilon=0.1, \alpha_m = {\cal P}_m = \chi_m=1$). As a first approach, we impose a value of the ejection index $\xi$ and vary the turbulence parameter $\alpha_P$ to understand the difference introduced by the turbulent pressure. As a second step, we explore the whole parameter set for $\xi$ and analyze how the turbulent pressure modifies the super-SM JED parameter space.


\begin{figure}
      \resizebox{0.95\hsize}{!}{\includegraphics{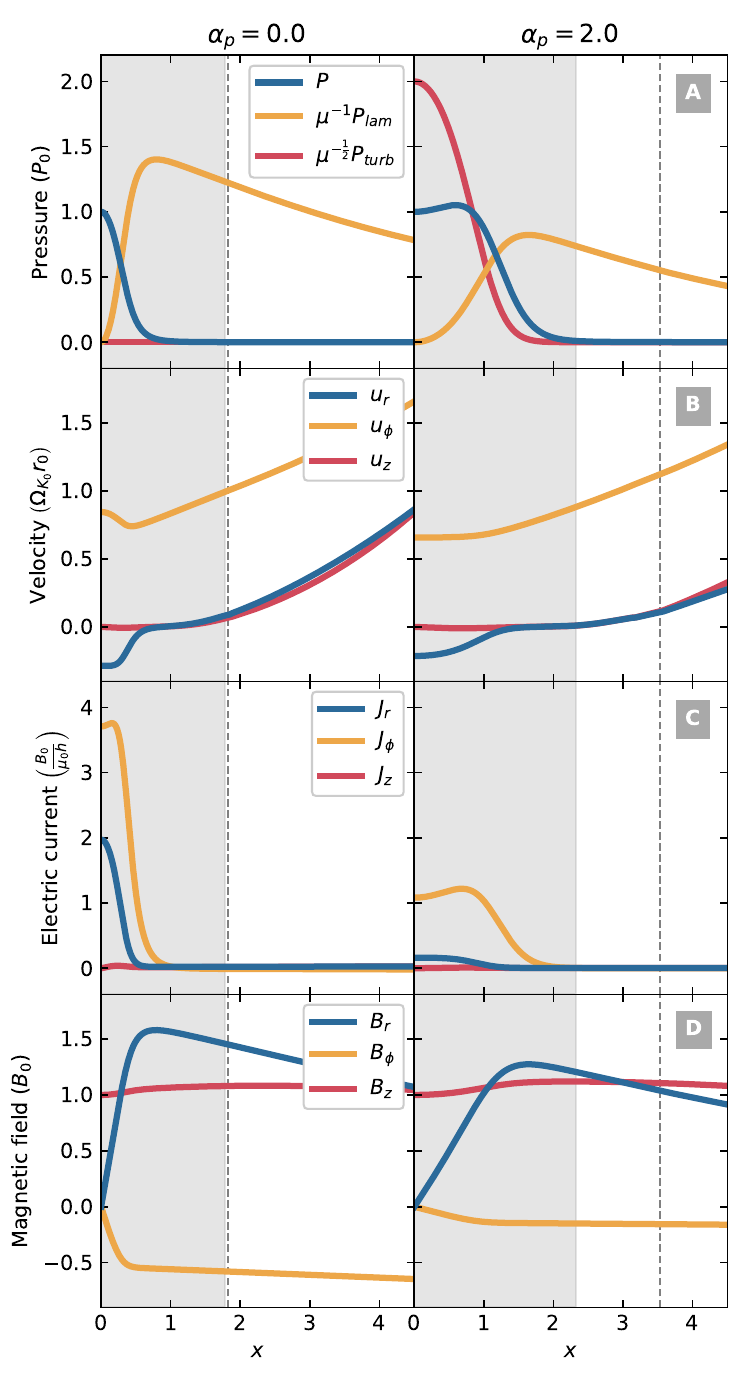}}
      \caption{Vertical profiles along $x = z/h$ of the two super-SM solutions shown in Fig.~\ref{fig:2D_disk}, without (left) and with (right) turbulent magnetic pressure. From top to bottom: gas $P$, laminar horizontal $P_{lam} = (B_r^2 + B_\phi^2)/2\mu_o$ and turbulent $P_{turb}$ magnetic pressures; velocity components $u_i$ (normalized to Keplerian speed  $\Omega_{K_0} r_0$); electric current densities $J_i$ (normalized to $B_{zo}/\mu_oh$); laminar magnetic field components $B_i$ (normalized to $B_{zo}$). The light gray area corresponds to the resistive MHD region and the vertical line corresponds to the position of the SM point. Note that $\alpha_P = 0$ requires $\mu = 0.6$, while  $\alpha_P = 2$ requires $\mu = 3.9$. See text for more details.}
      \label{fig:profiles_disk}
\end{figure}


\subsection{Comparative analysis}

Figure~\ref{fig:2D_disk} shows two super-SM solutions with the same ejection efficiency $\xi = 3 \times 10^{-3}$, one obtained without turbulent pressure ($\alpha_P = 0$, top panel) and corresponding to previously published solutions, and the other with turbulent pressure ($\alpha_P = 2$, bottom panel). At first glance, the main visible effect is to make the disk puffier, which makes perfect sense since turbulence provides additional support against vertical compression (due to both gravity and laminar magnetic field compression). However, $\alpha_P = 0$ requires $\mu= 0.6$ to get a trans-SM outflow, whereas $\alpha_P = 2$ requires $\mu = 3.9$. This is easily understood from the disk vertical balance: the stronger the laminar field (larger $\mu$), the stronger the vertical compression. Thus, imposing the same $\xi$ while introducing an extra, outwardly directed vertical force, leads to the natural increase of $\mu$. However, the impact of the turbulent pressure is more complex and affects all quantities.

Figure~\ref{fig:profiles_disk} shows the vertical profiles along $x = z/h(r)$ of various disk quantities. The left panels display the classical JED solution with $\alpha_P = 0$, while the right panels are for $\alpha_P = 2$. In both cases, the light gray area corresponds to the resistive MHD region, the white one to the ideal MHD regime where the flow reaches the super-SM speed. Clearly, introducing a turbulent pressure leads to a thicker disk, although the thermal scale height ($\epsilon = h/r$) remains the same: all disk quantities follow the turbulent pressure. We can understand this new disk structure from simple analytical considerations.
 

\begin{figure}
      \resizebox{\hsize}{!}{\includegraphics{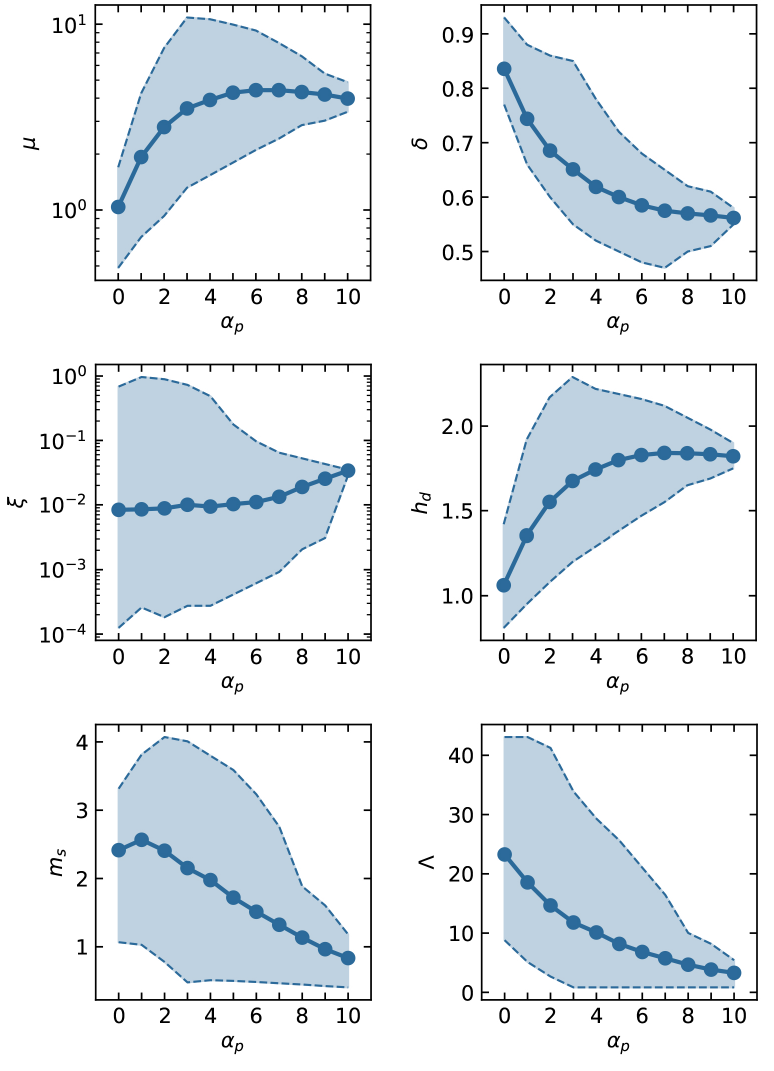}}
      \caption{Influence of $\alpha_P$ on the super-SM JED parameter space: disk magnetization $\mu$, deviation to Keplerian rotation $\delta$, disk ejection efficiency $\xi$, thickness expansion $h_d= z_{id}/h$, accretion sonic Mach number $m_s$,  and laminar (jets) to turbulent (viscous) torque ratio $\Lambda$. Colored areas correspond to regions accessible by super-SM solutions. The curves represent the average solution for a given $\alpha_P$. See text for more details.}
      \label{fig:alphap_disk}
\end{figure}

 
The fact that the disk becomes puffier can be seen directly on the vertical profile of the gas pressure $P$ (which is the same as the density profile, Fig~\ref{fig:profiles_disk}a). There is almost an inversion of the profile, which can be obtained directly by making a Taylor expansion of the vertical equilibrium Eq.~\ref{momentumconservation} near the disk midplane:
\begin{equation}
    \frac{\partial \ln P}{\partial x} \simeq - \frac{1 - \alpha_P \sqrt{\mu} x_{t}^{-2} + \mu (p^2 + q^2)}{1 + \alpha_P \sqrt{\mu}} x = -Ax \, ,
\label{eq:P}
\end{equation}
\noindent where $J_{ro}= q B_{zo}/(\mu_oh)$ and $J_{\phi o}= p B_{zo}/(\mu_oh)$ are respectively the radial and toroidal electric current densities flowing at the disk midplane. For $\alpha_P = 0$, only the thermal and laminar magnetic horizontal pressure gradients (the terms $1 + \mu (p^2 + q^2)$) play a role in the vertical equilibrium, so that the thermal pressure $P$ always decreases ($A > 0$). In that case, the isothermal disk scale height $h(r)$, defined as $P_o = \rho_o \Omega_K^2 h^2$, is indeed a good proxy for the actual disk vertical magnetostatic scale height \citep{ferr95}. This is no longer the case with turbulent pressure. For $\alpha_P = 2$, the turbulent magnetic pressure gradient $- \alpha_p \sqrt{\mu} x_{t}^{-2}$ balances the other two pressure gradients, so that the thermal pressure $P$ increases near the disk midplane ($A < 0$). The disk material is thus pushed toward the disk surface, making it thicker and more massive, as illustrated in Fig.~\ref{fig:2D_disk}.
 
Another direct consequence of this extra pressure can be seen in the disk rotation rate at the midplane (Fig.~\ref{fig:alphap_disk}). The deviation from the Keplerian rotation rate at the disk midplane is provided by the radial force balance 
\begin{equation}
    \delta = \frac{\Omega_o}{\Omega_{Ko}} \simeq \sqrt{1- \left( 1+\alpha_P \sqrt{\mu} \right) \left( \frac{5}{2} - \xi \right)\epsilon^2  - p \mu  \epsilon - \frac{m_s^2}{2} \epsilon^2 }  \, ,
    \label{deltaeq}
\end{equation}
\noindent where $m_s = -u_{ro}/C_{so} = - u_{ro}/(\Omega_{Ko} h)$ is the accretion sonic Mach number. In the RHS of this equation, the forces leading to a deviation are (going from left to right), the radial (thermal and turbulent) pressure gradient, the laminar magnetic tension and the radial acceleration. At near equipartition field, the turbulent radial pressure gradient becomes relevant and the disk is significantly sub-Keplerian ($\delta \sim 0.6$ here).
  
Cold jets are only possible if magnetic field lines at the disk surface are bent enough, namely if $B_r/B_z$ is greater than or on the order of unity (\citealt{blan82}, see Figs.~\ref{fig:2D_disk} and \ref{fig:profiles_disk}d). Since $B_r \simeq \mu_o \int_0^z J_\phi dz$, the field line inclination at the disk surface is roughly $B_r/B_z \simeq - \int (u_r/\nu_m) dz \sim p \int (f_{u_r}/m) dx$. As the disk inflates due to turbulence, the integral becomes larger, so $p$ must decrease. In other words, field lines become straighter at the midplane ($J_{\phi o}$ decreases), and $J_\phi$ is enhanced at higher altitudes. Although only moderate for the solution shown in Fig.~\ref{fig:profiles_disk}c, this trend can be significantly more pronounced for other solutions.

Correspondingly, the midplane accretion speed ($m_s$) decreases, which implies that the total torque acting on the disk must also decrease despite an increase in the disk magnetization $\mu$ to maintain the same ejection rate $\xi$. Accretion occurs due to the presence of a laminar (jets) torque and a turbulent (viscous) torque, namely
\begin{equation}
    m_s = \frac{2 q \mu}{\delta} + \alpha_v \epsilon \, . 
    \label{eqms}
\end{equation}
The ratio of the jet torques to the viscous torque scales as $\Lambda \sim 2/(\epsilon \alpha_m {\cal P}_m )$ (Eq.~35 in \citealt{jacq19}) and the viscous torque is therefore always negligible within our fiducial parameter set. In order for the midplane accretion speed to decrease, the jet torque $-J_{ro} B_{zo}$ must decrease despite an increase in $B_{zo}$. This dramatic decrease in $J_{ro}$ (straighter field lines) is clearly seen in Fig.~\ref{fig:profiles_disk}c for $\alpha_P=2$.
     
The evolution of the radial electric current density $J_r$ (Eq. \ref{inductioneq}) writes
  \begin{equation}
   \eta'_m J_r \simeq \eta'_{mo} J_{ro} + r \int_0^z B_r \frac{\partial \Omega}{\partial r} dz \, .
   \label{eq:Jr}
\end{equation}
\noindent The first term on the RHS of this equation results from the e.m.f induced by the rotation of a conductive disk (of finite resistivity) amidst a large-scale vertical field (Barlow's wheel effect). The second term is the disk feedback, always present whenever there is a velocity shear. Magneto-centrifugally driven jets are only possible if the laminar (jet) torque accelerates the ejected material, namely if $F_\phi \simeq - J_r B_z$ becomes positive at the disk surface \citep{ferr95}. This occurs only if $J_r$ decreases vertically on the same scale, a situation achieved whenever the shear-induced term in Eq.(\ref{eq:Jr}) balances the Barlow term within the disk. 

In a classical JED with $\alpha_P=0$, the vertical decrease of $J_r$ on a scale height (which goes along a vertical decrease of $\Omega$), leads to an almost linear growth of the toroidal field that achieves a value $B_\phi^+ = - \mu_o \int J_r dz \sim - \mu_o J_{ro}h $ at the disk surface, namely $q \sim 1$. Therefore all three magnetic field components are comparable at the disk surface (Fig.~\ref{fig:profiles_disk}d, left). But when turbulent pressure is present, rotation becomes smaller, the poloidal field lines are straighter and $B_r$ remains smaller inside the disk, two effects that heavily reduce the shear-induced term. This pushes further up the point in the disk where acceleration takes place ($F_\phi > 0$) and may even lead also to some profile inversion (barely visible in the right panel of Fig.~\ref{fig:profiles_disk}d, but representative of these turbulent solutions). Indeed, since $\eta'_m J_r \sim \eta'_{mo} J_{ro} $, a decreasing magnetic diffusivity leads to an increase of the radial current density at higher altitudes. The general solution, which is self-consistent with the disk vertical equilibrium, is therefore to reduce the midplane current density $J_{ro}$, with a tendency to shift the path of the radial electric current to both disk surfaces. This reduction has far-reaching consequences because it inhibits the generation of the toroidal magnetic field $B_{\phi}$, which is essential since it directly contributes to the accretion (torque) and ejection power (MHD Poynting flux).

\subsection{SM-parameter space}

We now study how the turbulent magnetic pressure modifies the super-SM parameter space of JEDs with our fiducial parameter set ($\epsilon=0.1, \alpha_m = {\cal P}_m = \chi_m=1$). For each value of $\alpha_P$ varying from 0 to 10, we explore all possible values of $\xi$ and derive the value $\mu$ that allows the outflowing plasma to become super-SM.

It is important to note that no Alfv\'enic constraint has been used yet so that the field line bending (parameter $p$) is freely imposed. In practice, we also explore all possible values of $p$ for each $\xi$. As a consequence, the number of solutions found for each $\xi$ is variable for each $\alpha_P$. This allows us to compute an "average" Super-SM solution for each $\alpha_P$, which simply shows the average value for that quantity (based on the number of solutions found). The influence of  $\alpha_P$ on the resulting parameter space is shown in Fig.~\ref{fig:alphap_disk}. The shaded zones for each quantity correspond to regions that are accessible to super-SM solutions, regardless of their number. The mean solution is then represented by a point which, as $\alpha_P$ increases, gives an indication of how the JED structure adapts to $\alpha_P$.

The actual size of the accretion disk can be very precisely defined as the altitude where the radial velocity is zero, namely the transition from inward (accretion) to outward (ejection) motion. Such a transition occurs in the resistive MHD region, and only further out does the flow transit to ideal MHD and becomes super-SM. The central right panel therefore shows the dependence of the actual disk size $h_d= z_{id}/h$ as a function of $\alpha_P$: clearly, the disk becomes puffier with increasing turbulent magnetic pressure, by almost a factor 2 when $\alpha_P = 10$. The reason for this has been discussed previously and is related to the fact that this turbulent pressure provides additional vertical support against gravity and magnetic (laminar) compression. Note that the range where super-SM solutions are found becomes very narrow beyond $\alpha_P\sim 7$. This is an indication that finding super-A solutions at large turbulent pressures will hardly be feasible (see next section).

An unsuspected consequence of the disk becoming thicker with $\alpha_P$ is the fact that super-SM outflows also require increasingly large disk magnetizations $\mu$ (Fig.~\ref{fig:alphap_disk}). This is actually an implication of our methodology which focuses only on non-oscillating solutions within the disk. To forbid spatial oscillations, one must require that the MRI wavelength be comparable to the disk vertical height, which translates into having comparable Alfv\'en and sound speeds. But our prescription of compressible MHD turbulence leads to a modification of the sound speed or, in other words, to an increase of the actual size ($h_d>1$) of the disk. Thus, to recover a non-oscillating solution one needs to balance $V_A \sim \tilde C_s$ and to redefine an MRI-related magnetization parameter 
  \begin{equation}
    \mu_{MRI} = \frac{V_A^2}{\tilde C_s^2} = \frac{\mu}{1+\alpha_p \sqrt{\mu}} \, .
    \label{muMRIequation}
\end{equation}
\noindent Thus, requiring to deal only with non-oscillating solutions, namely keeping $\mu_{MRI}$ constant while increasing $\alpha_P$, will necessarily lead to also increase $\mu$ as well. 


\begin{figure}
      \centering
      \resizebox{0.8\hsize}{!}{\includegraphics{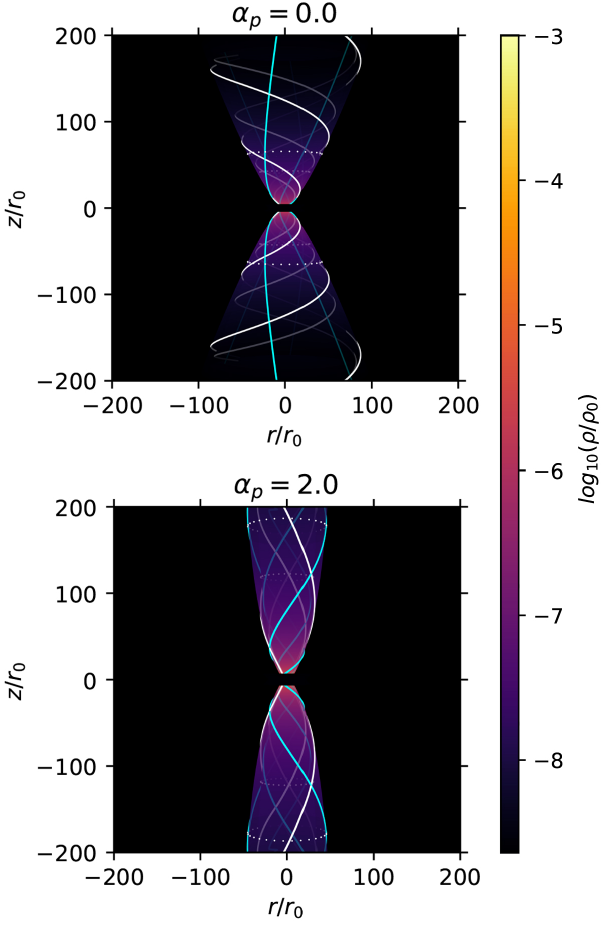}}
      \caption{3D view of the JED solutions shown in Fig.~\ref{fig:2D_disk} with the same ejection index ($\xi = 3 \times 10^{-3}$), obtained with our fiducial set ($\epsilon = 0.1, \alpha_m = \mathcal{P}_m = \chi_m=1$), without ($\alpha_P = 0, \mu = 0.6$) and with ($\alpha_P = 2, \mu = 3.9$) turbulent magnetic pressure. The color represents the logarithm of the density normalized to its midplane value, the white lines show the magnetic field lines and the blue lines show the streamlines along the same magnetic surfaces, anchored at $r_o = 2$ and $r_o = 3$. The white dotted lines show the position of the Alfv\'en point on these surfaces.}
      \label{fig:3D_jet}
\end{figure}


However, this trend reaches a limit around $\alpha_P\sim 4$, where $\mu$ levels off. This is because when $\mu$ becomes too large, not only the vertical balance is compromised (overwhelming laminar magnetic compression), but also the radial one (see Eq.~\ref{deltaeq}). The top right panel in Fig.~\ref{fig:alphap_disk} shows indeed that the deviation $\delta= \Omega_o/\Omega_K$ from the Keplerian rotation decreases dramatically reaching almost $\delta\sim 0.5$. Since gravity does not change, lowering the plasma rotation rate is clearly not convenient for launching cold outflows. This value $\alpha_P \sim 4$ marks therefore the point where the super-SM region starts to shrink.

The effect of turbulent magnetic pressure on the disk ejection efficiency $\xi$ is not obvious. At first sight, one would hope to enhance $\xi$ thanks to this extra outward push. This effect can indeed be seen in the central left panel in Fig.~\ref{fig:alphap_disk}: the minimum value for $\xi$ does indeed increase with  $\alpha_P$. Close to $10^{-4}$ for $\alpha_P=0$, it cannot be less than a few $10^{-2}$ for $\alpha_P=10$. However, the maximum value of $\xi$ that allows for super-SM outflows also decreases: this is because the disk is rotating slower and the magnetic laminar compression becomes too strong (it scales as $\mu$, while turbulence scales as $\mu^{1/2}$). Overall, the average super-SM JED solution (solid line) does not seem to see its ejection efficiency deviating significantly from the classical ($\alpha_P=0$) value, namely $\xi \sim 10^{-2}$ for isothermal flows.
 
Another counterintuitive effect of the turbulent magnetic pressure is to reduce the midplane accretion velocity, despite the increase in disk magnetization $\mu$. The bottom left panel in Fig.~\ref{fig:alphap_disk} indeed shows a decreasing sonic Mach number $m_s$ with increasing $\alpha_P$ (although it remains larger than, or on the order of, unity). This is due to the decreasing influence of the jet torque wrt the turbulent (viscous) torque. The bottom right panel clearly shows a steady decrease in the ratio $\Lambda$ of these two torques with $\alpha_P$. This is a direct consequence of the turbulent JED becoming unable to generate a strong toroidal magnetic field at the disk surface. As discussed previously, not only is the radial electric current density $J_{ro}$ (parameter $q$ in Eq.~\ref{eqms}) flowing in the midplane to be lowered in the presence of turbulent pressure, but the vertically decreasing magnetic diffusivity does not allow for a large toroidal field at the disk surface. As a consequence, jets play a less and less dominant role in accretion.\\
 
To conclude, our analysis of the inclusion of a turbulent magnetic pressure in JEDs shows that the disk becomes puffier, more sub-Keplerian, and less conductive, with field lines more vertical inside the disk. The generation of magnetic bending ($B_r$) and shear ($B_\phi$) occurs mostly at the (more elevated) disk surface, with no significant change in the disk ejection efficiency. Overall, a magnetic turbulent pressure reduces the impact of the jets on the accretion.

\section{Impact on the disk+jets: Super-A solutions}
\label{sec:super-A}


\begin{figure}
      \resizebox{\hsize}{!}{\includegraphics{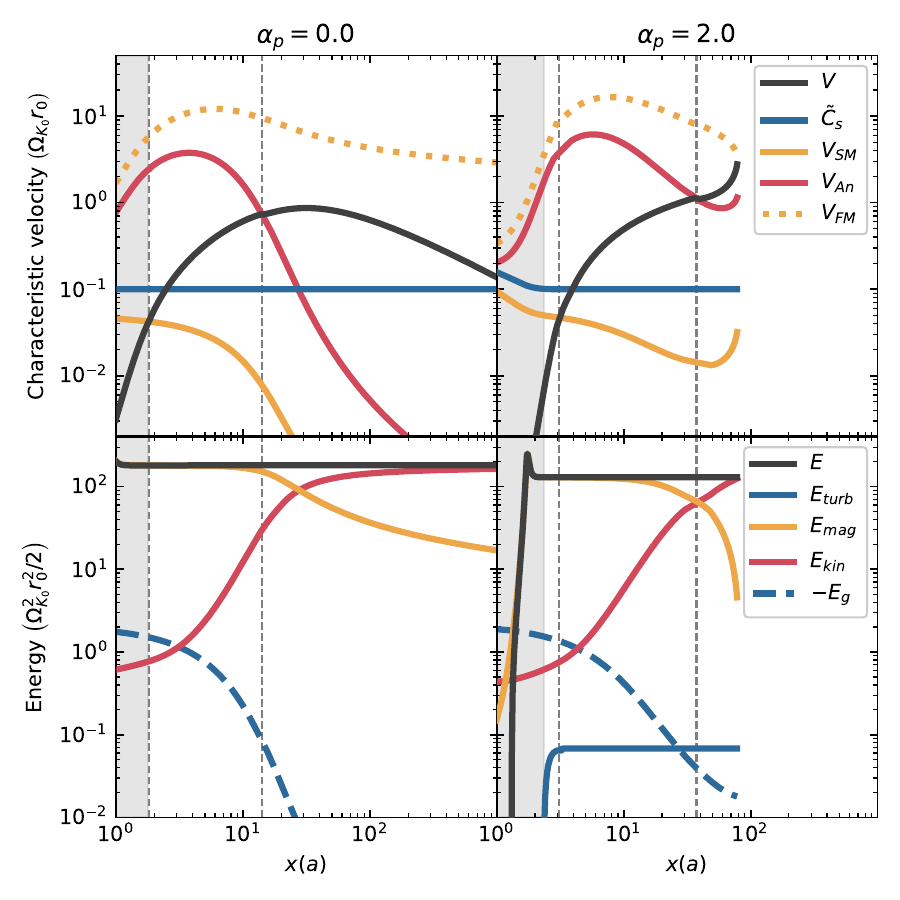}}
      \caption{Vertical profiles along $x = z/h(a)$ of the two super-A solutions shown in Fig.~\ref{fig:3D_jet}, without (left) and with (right) turbulent magnetic pressure. Top: characteristic velocities normalized to the Keplerian velocity at the disk midplane (flow speed $V$, modified sound speed $\tilde C_s$, slow and fast magnetosonic speeds $V_{S\!M}, V_{F\!M}$ and Alfv\'en speed $V_{An}$). Bottom: contributions to the Bernoulli invariant E normalized to $\Omega_{Ko}^2 r_o^2/2.$. The light gray area corresponds to the resistive MHD region and the vertical lines corresponds to the position of the SM and A point respectively. See text for details.}
      \label{fig:profiles_jet}
\end{figure}


\subsection{Comparative analysis} 

Figure~\ref{fig:3D_jet} shows a 3D view of the two JED solutions discussed previously, obtained with our fiducial parameter set ($\epsilon = 0.1, \alpha_m = {\cal P}_m = \chi_m = 1$). Our goal is to assess how the presence of a turbulent magnetic pressure impacts the dynamics of the outflowing super-A jet and, in turn, affects the underlying disk. Since both solutions have the same ejection efficiency ($\xi = 3 \times 10^{-3}$), they should have roughly the same Alfv\'en radius $r_A$, since it depends mostly on $\xi$ (F97, see also below). But a simple look at the figure reveals that the turbulent JED launches much less open (i.e., more collimated) jets with an Alfv\'en point located at a higher altitude $z_A$. Also, in agreement with the super-SM analysis, the magnetic field lines appear much less twisted, a sign that the jets may eventually become exhausted of any poloidal electric current, leading to a vanishing $B_\phi$.

Figure~\ref{fig:profiles_jet} shows the profiles of several quantities along a magnetic surface (constant $a$) for the two previous super-A solutions. The top row displays the various characteristic velocities normalized to the Keplerian speed at the anchoring radius $r_o$: the flow speed in the self-similar direction $V=\Vec{u}_p \cdot \vec{n}$, the modified sound speed and the modified A, SM, and FM phase speeds defined by Eq.~\ref{eqvsm} (see also \citealt{ferr95}).
Due to the Gaussian decrease of the turbulence, the influence of the turbulent pressure on the sound speed rapidly decreases as the plasma is lifted out of the disk. Thus, once in ideal MHD, only the real thermal energy remains ($\tilde{C}_s = C_s$) and the sound speed remains constant along the field lines, in agreement with our assumption. This can also be seen in the Bernoulli integral (bottom panels, Fig.~\ref{fig:profiles_jet}): the turbulent energy content $E_{turb}$ remains completely negligible. In the isothermal case, the enthalpy is also always negligible and the jets are therefore cold, the dominant energy being initially stored in the magnetic field $E_{mag}$ and progressively transferred to the kinetic energy $E_{kin}$ of the plasma.


\begin{figure}
      \resizebox{\hsize}{!}{\includegraphics{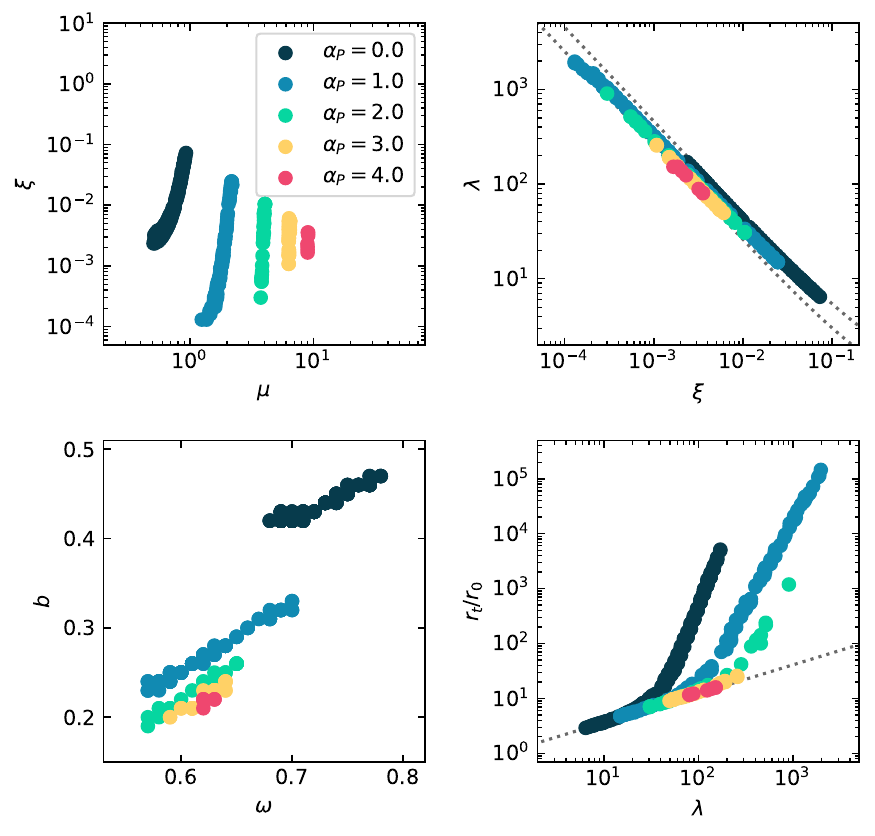}}
      \caption{Influence of the magnetic turbulent pressure $\alpha_P$ on super-A solutions obtained with our fiducial set. Top left: JED parameter space $\xi(\mu)$. Top right: magnetic lever arm $\lambda$ as function of $\xi$. The dotted lines correspond to $\lambda= a[1 + 1/(2 \xi)]$ with $a =.9$ and $a = 0.5$ Bottom left: fraction $b$ of the released accretion power transferred to the two jets as function of the magnetic surface rotation $\omega$ anchored at a radius $r_o$. Bottom right: maximum cylindrical radius $r_t/r_o$ reached before recollimation, as function of $\lambda$. The dotted line is a fit $r_t/r_o = 1.08\lambda^{0.526}$.}
      \label{fig:edp_jet}
\end{figure}


\subsection{A-parameter space}

The top left panel in Fig.~\ref{fig:edp_jet} shows how the usual JED parameter space $\xi(\mu)$ evolves with $\alpha_P$. For each value of $\alpha_P$, we recover the characteristic almost vertical behavior, namely the ability to achieve a wide range of ejection efficiencies $\xi$ in a very narrow interval of $\mu$. Three important features emerge immediately as $\alpha_P$ increases: (1) $\mu$ must be increasingly large, (2) $\xi$ decreases drastically, (3) until no super-A solution can be found beyond $\alpha_P=4$.
The fact that $\mu$ must increase with $\alpha_P$ is a consequence of keeping the MRI wavelength of nearly the same size as the puffed disk (as described by Eq.~\ref{muMRIequation}). It is therefore a consequence of the disk vertical equilibrium, dealt with the SM critical point. However, it is interesting to note the existence of a "sweet spot" for $\alpha_P=1$, associated with the largest interval in $\xi$. But, globally, the super-A parameter space, which is a subset of the previous SM parameter space, shrinks with $\alpha_P$, with a global decrease of the maximum ejection efficiency $\xi_{max}$. Such a decrease in $\xi_{max}$ was already present in the SM parameter space (see Fig.~\ref{fig:alphap_disk}), but it is greatly amplified by the Alfv\'enic constraint.

This can be understood by the fact that the disk rotation rate $\delta$ decreases with $\alpha_P$, resulting in a global decrease in the rotation rate 
\begin{equation}
\omega = \frac{\Omega_*(a)}{\Omega_{Ko}}
\end{equation}
of the magnetic surfaces (bottom left panel in Fig.~\ref{fig:edp_jet}). Since the magneto-centrifugal acceleration is the only means for launching our isothermal (cold) outflows, less and less disk material can be accelerated to super-A speeds as $\alpha_P$ increases.
This interpretation is validated by the neat correlation $b(\omega)$ for all $\alpha_P$, where $b= 2 P_{jet}/P_{acc}$ is the ratio of the total power (kinetic, thermal, potential, and magnetic) carried away by the two jets to the released accretion power. The reason is simply that the main energy content is magnetic and that it scales directly with $\Omega_*$, namely $P_{jet} = \int \rho \Vec{u}_p E\cdot \Vec{dS}\simeq  \int \Vec{S}_{MHD}\cdot \Vec{dS}$, where the MHD Poynting flux is $\Vec{S}_{MHD} = - \Omega_* r B_{\phi} \Vec{B}_p/\mu_0$. As guessed from the study of the SM parameter space, magnetic turbulence quite significantly reduces the amount of laminar magnetic energy feeding the two jets.


\begin{figure}
      \resizebox{\hsize}{!}{\includegraphics{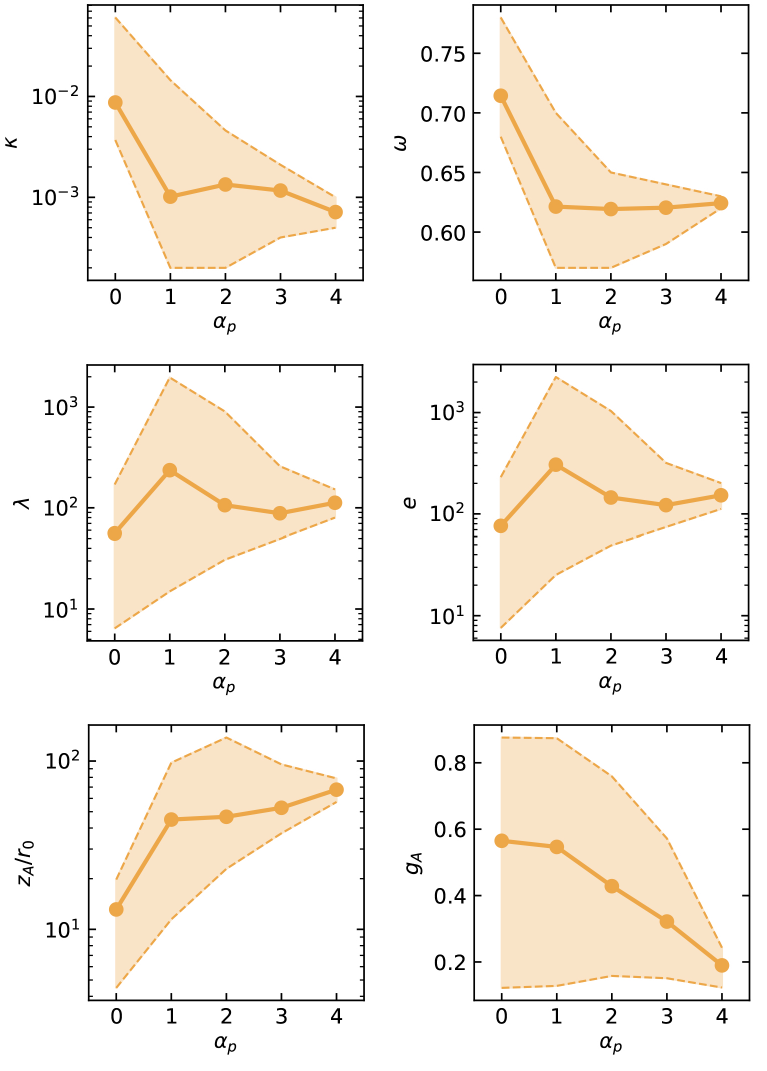}}
      \caption{Influence of $\alpha_P$ on the super-A JED parameter space: jet mass load $\kappa$, rotation rate $\delta$ of the magnetic surface, magnetic lever arm $\lambda$, total specific energy $e$ carried along the magnetic surface, altitude $z_A/r_o$ of the Alfv\'en point, and the electric current $g_A$ still available at that point. Colored areas correspond to regions accessible by super-A solutions. The curves represent the average solution for a given $\alpha_P$. See text for more details.}
      \label{fig:alphap_jet}
\end{figure}


The total specific angular momentum $L(a)$ carried along a magnetic surface (Eq. \ref{L}), normalized by the Keplerian angular momentum at the anchoring radius $r_o$, provides the so-called magnetic lever arm parameter $\lambda$ introduced by \citet{blan82}. This important jet quantity is actually related to the disk parameters
\begin{equation}
 \lambda = \frac{L (a)}{\Omega_{Ko} r_o^2} \simeq \omega + \frac{1}{2 \xi} \frac{\Lambda}{1+ \Lambda} \, ,
\end{equation}
which is a generalization of the classical relation $\lambda \simeq 1 + 1/2\xi$ (F97). Note that most JED solutions have $\Lambda >> 1$. 
The top right panel in Fig.~\ref{fig:edp_jet} shows how the dependency $\lambda(\xi)$ evolves with $\alpha_P$. The two bracketing dotted lines are the curves $\lambda = a (1 + 1/2\xi)$, with $a=0.5$ and $a=0.9$. They show that the analytical link is not broken and that any measure of $\lambda$ can give a fair estimate of $\xi$. The main difficulty, however, is the decrease of the disk rotation rate (hence $\omega$) as $\alpha_P$ increases, leading to a decrease in $\lambda$ at constant $\xi$.

Nevertheless, turbulent magnetic pressure in the disk leads to a dramatic difference in the jet collimation properties, as illustrated in the bottom right panel in  Fig.~\ref{fig:edp_jet}. This plot displays the jet widening, namely the ratio $r_t/r_o$ where $r_t$ is the maximum cylindrical radius reached by a magnetic surface anchored at $r_o$ before recollimating toward the axis (F97), as a function of the magnetic lever arm $\lambda$ and for different values of the turbulent magnetic pressure parameter $\alpha_P$. The gray dotted line is a numerical fit $r_t/r_o= 1.08\lambda^{0.526}$, namely $r_t\simeq r_A$. Here we recover the existence of the sweet spot at $\alpha_P=1$, which allows us to obtain the widest types of behavior, namely from solutions no larger than $r_A$ to solutions opening up to $10^5 r_o$.
But it can be readily seen that as  $\alpha_P$ increases, there are fewer and fewer solutions that propagate much further than the Alfven point. At $\alpha_P= 4$, all solutions found merely become super-A before undergoing recollimation. Such solutions, termed "current-free" in F97, become super-A at the expense of almost all the available electric current flowing out of the disk.

Figure~\ref{fig:alphap_jet} shows the evolution of several jet parameters as a function of the turbulent magnetic pressure intensity $\alpha_P$. The colored areas correspond to regions accessible by super-A solutions, while the curves represent the evolution of the mean solution obtained by averaging the values of all the solutions found for a given $\alpha_P$. These figures allow us to better understand the influence of the disk turbulent pressure on the jet parameters and dynamics.

The jet mass load, defined as \citep{blan82}
\begin{equation}
\kappa = \eta(a) \frac{\Omega_{Ko} r_o} {B_{zo}} \simeq  \frac{\xi m_s}{\mu} = \xi \alpha_m p \mu^{-1/2} \, , \label{kappa}
\end{equation}
is an MHD invariant that can be related to the disk quantities using $\xi$. The value of $p$, which is a measure of the toroidal electric current density flowing at the disk midplane and loosely related to the field lines bending at the disk surface, is determined by the smooth crossing of the Alfv\'en critical point. The larger $\kappa$ the more kinetically dominated the outflow is. It turns out that for JED (namely strongly magnetized) solutions $\kappa \sim \xi$ (F97, \citealt{cass00a}), so we show below only $\xi$. The top left panel in Fig.~\ref{fig:alphap_jet} proves that only very tenuous, magnetically dominated cold outflows can become super-A. This is amplified by both the increase in $\mu$ and the corresponding decrease of $\xi$ as $\alpha_P$ increases.
 
As shown above, this is mostly due to the dramatic decrease in the rotation of the magnetic surface $\omega$ with $\alpha_P$. Indeed, as $\mu$ increases, the midplane plasma rotation $\delta$ decreases and, despite the MHD acceleration starting at higher altitudes, the plasma hardly becomes super-Keplerian \citep{ferr95}. As a consequence, the centrifugal term appearing in the radial momentum equation does not overcome gravity and only the radial component of the laminar magnetic force $F_r = J_\phi B_z - J_zB_\phi$ must become capable of driving an outwardly directed flow. While super-SM outflows are possible at $\alpha_P=10$ with $\delta\simeq 0.55$, no super-A solution is found with $\alpha_P > 4$ (for our fiducial parameter set), the lowest value being $\omega \simeq 0.63$ (corresponding to $\delta \simeq 0.63$ as well).


For cold outflows, and neglecting any additional energy $E_{turb}$ associated with the turbulent magnetic pressure, Eq.~\ref{E} leads to a dimensionless Bernoulli invariant
\begin{equation}
e = \frac{2 E(a)}{\Omega_{Ko}^2 r_o^2 }  \simeq \omega \lambda - 2 + \frac{\omega^2}{2} \, . \label{e}
\end{equation}
The exact value $e(a)$ is shown at the central right panel in Fig.~\ref{fig:alphap_jet}. Not surprisingly for cold outflows, it fulfills the above equation and closely follows the evolution of $\lambda$, which is itself directly related to the ejection efficiency $\xi$.
Our outflows from turbulent disks are not only gravitationally unbound ($e > 0$), but they are also highly magnetized ($e >> 1$). The ratio of the (laminar) MHD Poynting flux to the matter-energy flux at the SM point writes 
\begin{equation}
    \sigma_{S\!M} = \left.\frac{- \Omega_* r B_{\phi} B_p}{\left(\frac{u^2}{2} + H\right) \mu_0 \rho u_p}\right\vert_{S\!M} \simeq 2\omega(\lambda - 1)
\end{equation}
for cold outflows \citep{jacq19}, and is therefore much larger than unity. One might naively expect such highly magnetized jets to propagate far from the disk. However, this is clearly not the case, as they do not extend much beyond the Alfvén point, as shown in the bottom left panel of Fig.~\ref{fig:alphap_jet}.


\begin{figure}
      \resizebox{\hsize}{!}{\includegraphics{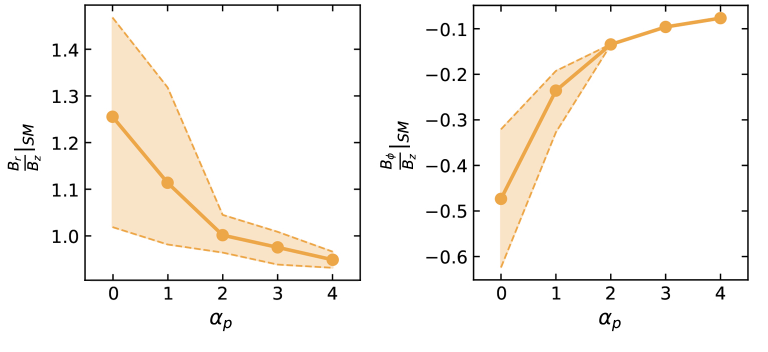}}
      \caption{Influence of $\alpha_P$ on the laminar magnetic field components at the SM critical point for super-A solutions obtained with our fiducial parameter set. Left: ratio $B_r/B_z$ showing the bending. Right: ratio $B_\phi/B_z$ showing the magnetic shear. Colored areas correspond to regions accessible by super-A solutions. The curves represent the mean solution for a given $\alpha_P$.}
      \label{fig:Bcomponents}
\end{figure}


This can be understood by looking at how the altitude $z_A/r_o$ of the Alfv\'en point evolves with $\alpha_P$ (Fig.~\ref{fig:alphap_jet}). As the turbulent magnetic pressure increases, the disk becomes puffier, and its magnetization $\mu$ increases. This implies a larger Alfv\'en velocity and thus would require a longer acceleration length to allow the jet poloidal speed to reach the Alfv\'en velocity. But a look at the SM parameter space (hence the disk vertical balance, Fig.~\ref{fig:alphap_disk}) shows that the ejection efficiency tends to remain nearly the same so that the cylindrical Alfven radius $r_A/r_o$ does not change much. As a consequence, increasing the acceleration distance can only be done by increasing $z_A/r_o$. Clearly, the interplay between the disk vertical balance (SM point) and the jet acceleration (A point) requires playing on both sides, namely increasing both $r_A$ (thus decreasing $\xi$) and $z_A$. For a given $\xi$, this ends up in a more collimated outflow emitted from a turbulent JED than from a nonturbulent JED, as illustrated in Fig.~\ref{fig:3D_jet}.

As the Alfv\'en point is pushed further away from the disk with increasing $\alpha_P$, the field lines at the disk surface become more and more straight (Fig.~\ref{fig:Bcomponents}, left panel). This appears to be a geometrical consequence, but it is due to the fact that in the sub-A region, the plasma is dominated by the magnetic field and in particular by the poloidal component. This magnetic tension always tends to close the magnetic surface, thus preventing plasma acceleration. For our fiducial parameter set, at $\alpha_P >4$ the magnetic bending is such that the energetic criterion for cold outflows (requiring $B_r/B_z >1$ at the disk surface, \citealt{blan82}) is no longer satisfied and no super-A solution can be found.

This decrease in the magnetic field bending at the disk surface goes along with a decrease in the toroidal magnetic field component (Fig.~\ref{fig:Bcomponents}). This is consistent with the decrease in the total energy feeding the jets (fraction $b$, (Fig.~\ref{fig:edp_jet}). Note that this behavior is due to the fact that we only consider JED structures with $\mu$ larger than unity. Indeed, for weakly magnetized accretion-ejection structures (WEDs) with $\mu << 1$, a large toroidal (laminar) magnetic field could be generated (see \citealt{jacq19}), but these solutions are beyond the scope of the present paper.

This tremendous decrease in magnetic shear at the disk surface goes along with a much lower acceleration efficiency, consistent with the increasing distance $z_A$ from the Alfv\'en point. As a proxy for this efficiency, one can use the function $g = 1 - \frac{\Omega}{\Omega_*}$, which measures the angular velocity drift between the plasma and the magnetic surface \citep{pell92}. It increases monotonically from $g<<1$ at the disk surface, while its value at the Alfv\'en point,
\begin{equation}
 g_A \simeq \frac{r_A B_{\phi, A}}{r_{S\!M} B_{\phi,S\!M}} \,
\end{equation}
determines the fate of the jet (F97). Indeed, $g_A$ is the ratio of the electric poloidal current $I=r B_\phi$ (enclosed in the magnetic surface) still flowing at the Alfv\'en point to the one provided at the disk surface (taken here as the SM point).
Solutions with $g_A > 1/2$ are current-carrying outflows where the acceleration to super-A speeds has been done with not much expense of the initial magnetic energy: at the Alfv\'en point, the magnetic field still has a large reservoir of angular momentum (and energy) to transfer back to the plasma. On the contrary, solutions with $g_A < 1/2$ have exhausted all the available magnetic energy just to reach super-A speeds.

While current-carrying jets open wide and propagate much farther out than the Alfv\'en point, current-free jets with low $g_A$ tend to recollimate immediately upon reaching this point. This is mostly due to the dominant effect of the poloidal magnetic tension (F97). Figure~\ref{fig:alphap_jet} shows that as $\alpha_P$ increases, $g_A$ decreases down to $\sim 0.2$, leading therefore mostly to current-free solutions at larger $\alpha_P$ for our fiducial parameter set.

\section{Parameter study}
\label{sec:param}

\subsection{Effect of the disk geometrical thickness}

The results described in the previous sections were obtained for a fiducial set ($\epsilon = 0.1, \alpha_m = {\cal P}_m = \chi_m = 1$). In this section, we analyze the effect of the magnetic turbulent pressure on isothermal JED solutions by exploring two other values of the disk aspect ratio:  $\epsilon = 0.01$, characteristic of an optically thick, geometrically thin disk and $\epsilon = 0.2$, characteristic of a slim/optically thick or geometrically thick, optically thin disk (see e.g., \citealt{marc18a}).  The other parameters are kept constant, namely  ($\alpha_m = {\cal P}_m = \chi_m = 1$). As before, we discuss first the impact of the turbulent magnetic pressure on the disk structure (SM constraint only) and then the constraint imposed to obtain outflows with super-A speeds (both SM and A constraints).


\begin{figure}
      \resizebox{\hsize}{!}{\includegraphics{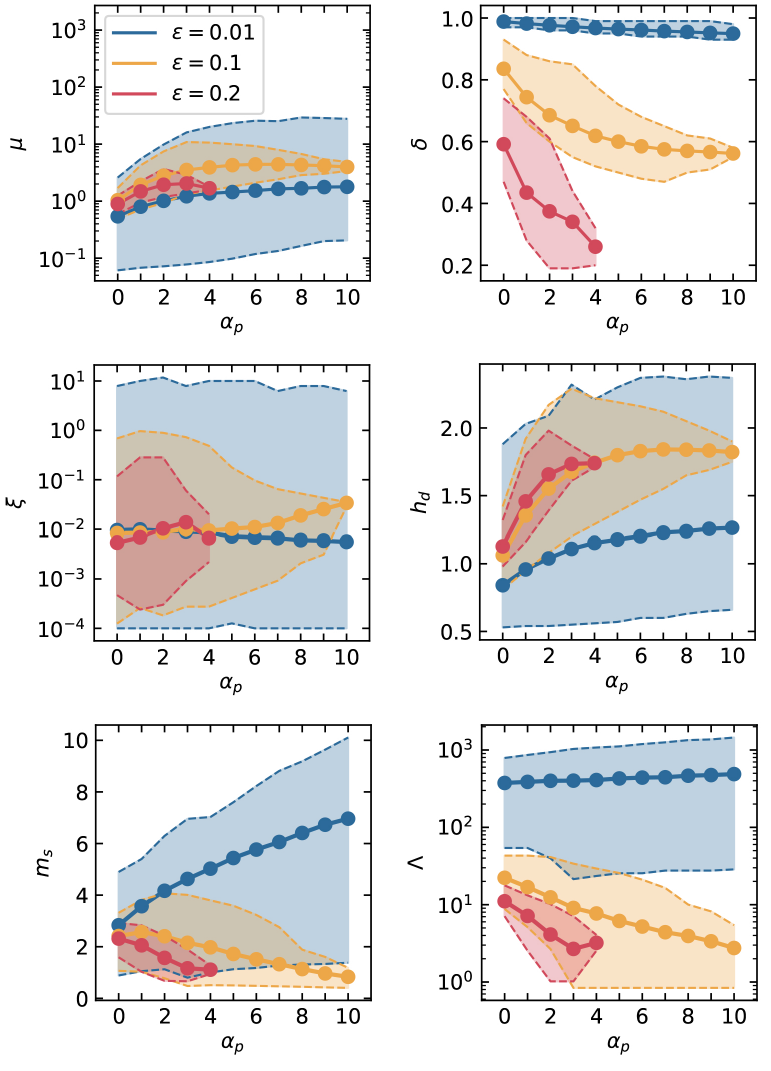}}
      \caption{Influence of $\alpha_P$ on the super-SM JED parameter space for various disk aspect ratios $\epsilon=h/r$ obtained with ($\alpha_m = {\cal P}_m = \chi_m = 1$): disk magnetization $\mu$, deviation to Keplerian rotation $\delta$, disk ejection efficiency $\xi$, thickness expansion $h_d= z_{id}/h$, accretion sonic Mach number $m_s$, and laminar (jets) to turbulent (viscous) torque ratio $\Lambda$. Colored areas correspond to regions accessible by super-SM solutions with different aspect ratios: $\epsilon = 0.01$ (blue),  $\epsilon = 0.1$ (yellow), $\epsilon = 0.2$ (red). The corresponding curves represent the average solution for a given $\alpha_P$.}
      \label{fig:alphap_epsilon_disk}
\end{figure}


Figure~\ref{fig:alphap_epsilon_disk} shows the evolution of several disk quantities as a function of the turbulent magnetic pressure intensity $\alpha_P$ for different disk geometrical thicknesses $\epsilon$. Colored areas correspond to regions accessible by super-SM solutions. The curves represent the evolution of the average solution obtained by averaging the values of all solutions found for a given $\alpha_P$.

As expected the magnetization $\mu$ first increases and then saturates in the same way regardless of the geometrical thickness $\epsilon$. This is because, as the disk inflates (i.e., $h_d= z_{id}/h$) due to the increasing importance of the magnetic turbulent pressure, the condition for the absence of MRI channel modes (spatial oscillations) remains fulfilled (Eq.~\ref{muMRIequation}).

This increase in the laminar magnetic field has tremendous consequences on the midplane disk rotation rate. In JEDs (strongly magnetized accretion disks), the dominant contribution to the deviation from the Keplerian rotation law is the laminar magnetic radial tension, which is proportional to $\mu\epsilon$ (Eq.~\ref{deltaeq}). But the radial gradient of the turbulent magnetic pressure provides a contribution that scales as $\alpha_P \mu^{1/2}\epsilon^2$, so it also becomes non-negligible at large $\mu$ and high $\alpha_P$. It is therefore not surprising that the deviation remains tiny in thin accretion disks, namely $\delta= \Omega_o/\Omega_{Ko} \sim 1$ at $\epsilon=0.01$ (blue, Fig.~\ref{fig:alphap_epsilon_disk}). But as the disk thickens,  the disk becomes more and more sub-Keplerian as $\alpha_P$ increases. For a thick disk with $\epsilon=0.2$ (red, Fig.~\ref{fig:alphap_epsilon_disk}), $\delta\sim 0.2$ at $\alpha_P=4$ and no super-SM solution can be found beyond this turbulence level. For even thicker disks ($\epsilon >0.2$) no super-SM (isothermal) solution can be found, even with $\alpha_P=0$ (F97, \citealt{cass00a}). The reason is quite simple: for cold jets, only magnetic driving allows acceleration so that whenever the plasma rotation becomes too low, the gravitational attraction always wins.

For the values of $\epsilon$ explored here, there is no significant effect on the values of the average disk ejection efficiency $\xi$ around $\sim 10^{-2}$ (central left panel of Fig.~\ref{fig:alphap_epsilon_disk}). However, the range of allowed $\xi$ for super-SM outflows changes significantly with $\epsilon$ and with $\alpha_P$. It is noteworthy to see how much larger the SM parameter space is for the thin disk, even with values $\xi > 1$ possible. We will come back to this point in Sect.~6.2.

The midplane accretion sonic Mach number $m_s = -u_{ro}/C_s = \alpha_m p \mu^{1/2}$ remains always larger than unity (supersonic accretion) for all values of the disk thickness. However, the dependency with the turbulent magnetic pressure is different for thin ($\epsilon = 0.01$) and for slim or thick disks ( $\epsilon=0.1$ and 0.2): while in the thin case $m_s$ increases with $\alpha_P$, it decreases in thicker disks (Fig.~\ref{fig:alphap_epsilon_disk}). For thick disks, which become even thicker as $\alpha_P$ increases, it has been argued that both radial and azimuthal electric currents are pushed toward the disk surface, leading to a correspondingly decrease of their value at the disk midplane (see Sect.~3.1). As a consequence, the equatorial field line curvature $p$ and jet torque become smaller leading to a decrease of $m_s$ with $\alpha_P$ (see also Eq.~\ref{eqms}). This also translates into structures where less and less disk angular momentum is carried away by the outflows (the torques ratio $\Lambda$ decreases).  This is no longer true in thin disks, where the turbulent magnetic pressure has a smaller impact on the disk final size (since $\Lambda$ scales with $1/\epsilon$). As a consequence, the current densities can remain mostly at the disk equatorial plane, which translate into magnetic configurations much more bent and sheared (both $p$ and $q$ of order unity). The increase of $m_s$ with $\alpha_P$ is therefore mostly a consequence of its dependence as $\mu^{1/2}$. We recover here the classical result obtained with $\alpha_P = 0$ that both the laminar (jet) magnetic torque and the turbulent (viscous) torque scale as $\mu^{1/2}$, so that their ratio $\Lambda$ does not evolve with $\alpha_P$ and is proportional to $\epsilon^{-1}$ \citep{cass00a,jacq19}.


\begin{figure}
      \resizebox{\hsize}{!}{\includegraphics{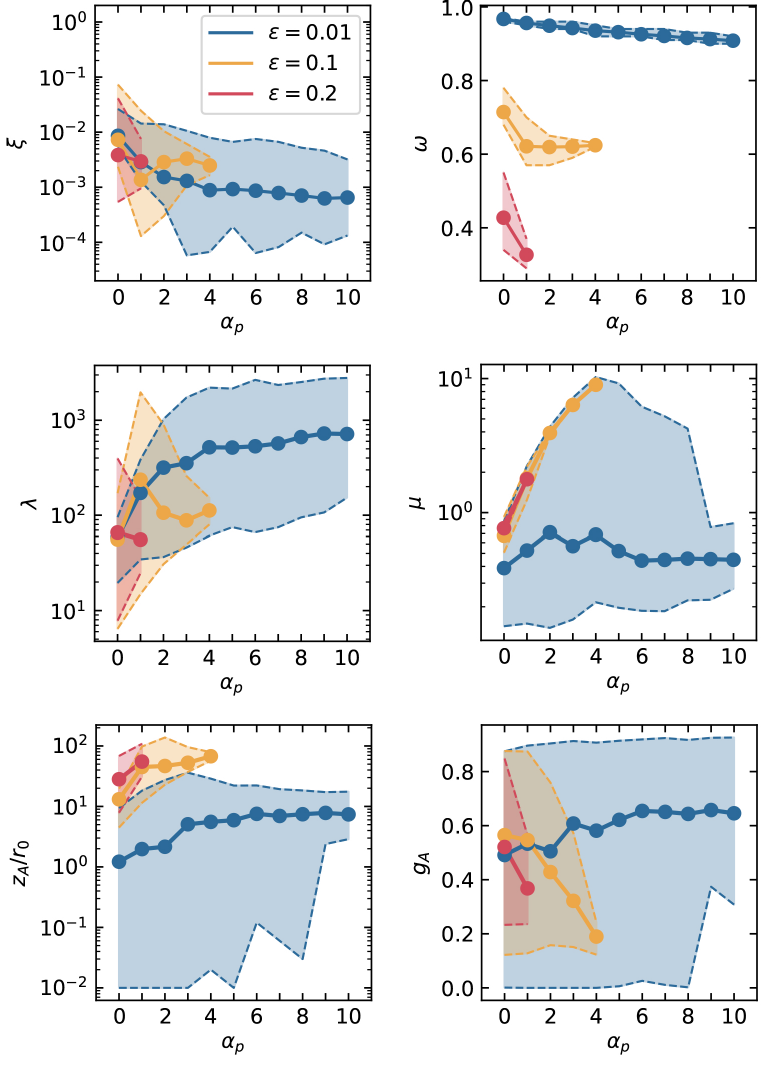}}
      \caption{Influence of $\alpha_P$ on the super-A JED parameter space for various disk aspect ratio $\epsilon = h/r$ obtained with ($\alpha_m = {\cal P}_m = \chi_m = 1$): disk ejection efficiency $\xi$, rotation rate $\omega$ of the magnetic surface, magnetic lever arm $\lambda$, disk magnetization $\mu$, altitude $z_A/r_o$ of the Alfv\'en point, and the electric current $g_A$ still available at that point. Colored areas correspond to regions accessible by super-A solutions with different aspect ratios: $\epsilon = 0.01$ (blue),  $\epsilon = 0.1$ (yellow), $\epsilon = 0.2$ (red). The corresponding curves represent the average solution for a given $\alpha_P$.}
      \label{fig:alphap_epsilon_jet}
\end{figure}


Figure~\ref{fig:alphap_epsilon_jet} shows the evolution with $\alpha_P$ of the parameter space for super-A solutions for different geometrical thicknesses $\epsilon$. As expected, it is much smaller than the SM parameter space studied previously. There is no solution beyond $\alpha_P = 1$ for $\epsilon = 0.2$ and beyond  $\alpha_P = 4$ for $\epsilon = 0.1$. The reason lies in the fact that the magnetic surface rotates much too slowly to magnetically drive cold (isothermal) outflows (see the $\omega$ panel), in agreement with our analysis of the disk rotation rate at the equatorial plane.

Despite the supplementary Alfv\'enic constraint, there is no better contrast in $\xi$ (or in the magnetic lever arm $\lambda$) achieved for different disk thicknesses. Indeed, the allowed ranges in $\xi$ stay nearly the same with a typical $\xi \sim 10^{-3}$. Note that the jet mass load $\kappa$ still always follows closely $\xi$.

The most striking feature is the very different behavior of the thin disk solution ($\epsilon=  0.01$, blue), as shown in the plots of the disk magnetization $\mu$, the altitude $z_A/r_o$ of the Alfv\'en point and the energy reservoir $g_A$. The typical thin disk solution (curve in the plot) maintains almost the same magnetization $\mu$ regardless of $\alpha_P$, in strong contrast to the other two thicker solutions. This can be understood by the fact that the slight increase in disk thickness due to $\alpha_P$ has not significantly changed the total pressure\footnote{In other words, the actual vertical scale height of the disk is not changed enough and there is no need to significantly increase $\mu$ to maintain almost the same "no oscillation" island as $\alpha_P = 0$ in Fig.~\ref{fig:edp_jet}. See also the discussion in \citet{jacq19}.}. As a consequence, the accretion-ejection can adapt its magnetic geometry while keeping roughly the same $\mu$. This has several consequences.
Instead of allowing more mass (larger $\xi$) to be ejected, the system favors solutions where the field lines become more and more bent at the disk surface as $\alpha_P$ increases. While a typical solution with $\alpha_P = 0$ has $B_r/B_z \sim 2$ at the SM point, it increases until $B_r/B_z \sim 5$ for  $\alpha_P = 10$. Magneto-centrifugal acceleration is thus enhanced, and typical outflows reach the Alfv\'en point at about the same location (see bottom left panel in Fig.~\ref{fig:alphap_epsilon_jet}), but without expending much of the initial magnetic energy. In contrast to thicker disks with $\epsilon = 0.1$ and $0.2$, whose jets' remaining energy $g_A$ at the Alfv\'en point decreases with $\alpha_P$, jets from a thin disk with $\epsilon = 0.01$ see their energy increase and remain larger than 1/2.

\subsection{Effect of the magnetic diffusivity} 

We now analyze how the turbulence level parameter $\alpha_m$, which determines both anomalous magnetic diffusivities and viscosity, affects the JED response to the turbulent magnetic pressure $\alpha_P$. Since classical JEDs are always dominated by the laminar jets torque, we will mostly interpret the role played by $\alpha_m$ as a modification of the disk magnetic diffusivities. In this context, it has already been shown that $\alpha_m$ must be large enough (of order unity) to build up a significant toroidal field at the disk surface (F97, \citealt{jacq19}). For this reason, we restrict our study to two other values $\alpha_m = 0.5$ and $\alpha_m=3$, while keeping the same set ($\epsilon=0.1, {\cal P}_m = \chi_m = 1$).


\begin{figure}
      \resizebox{\hsize}{!}{\includegraphics{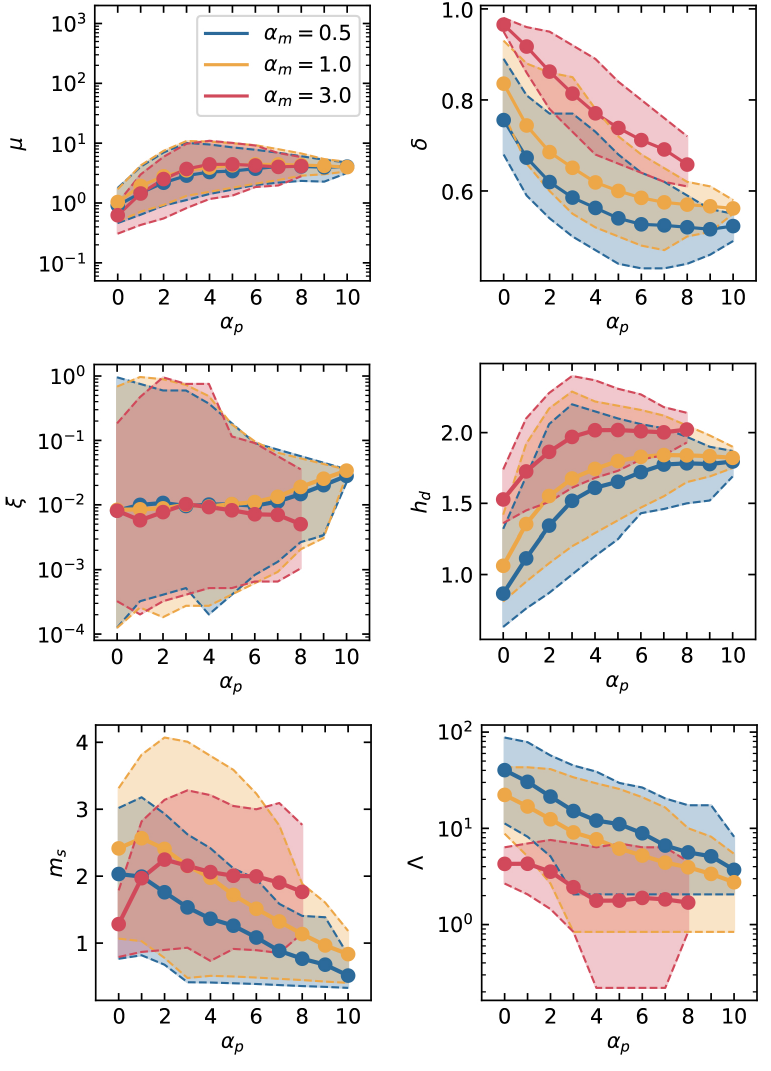}}
      \caption{Influence of $\alpha_P$ on the super-SM JED parameter space for various magnetic diffusivity levels $\alpha_m$ and ($\epsilon = 0.1, {\cal P}_m = \chi_m = 1$): disk magnetization $\mu$, deviation to Keplerian rotation $\delta$, disk ejection efficiency $\xi$, thickness expansion $h_d= z_{id}/h$, accretion sonic Mach number $m_s$, and laminar (jets) to turbulent (viscous) torque ratio $\Lambda$. Colored areas correspond to regions accessible by super-SM solutions with different magnetic diffusivity levels: $\alpha_m = 0.5$ (blue),  $\alpha_m = 1$ (yellow),  $\alpha_m = 3$ (red). The corresponding curves represent the average solution for a given $\alpha_P$.}
      \label{fig:alphap_alpham_disk}
\end{figure}


The parameter space obtained only with the SM constraint is displayed in Fig.~\ref{fig:alphap_alpham_disk} for the three values of $\alpha_m$. There is no significant modification introduced by $\alpha_m$ on the necessary increase of the disk magnetization $\mu$ as a response to an increase of the magnetic turbulent pressure $\alpha_P$ (same reason based on $\mu_{MRI}$). The same lack of strong influence can be seen in the allowed ranges of the disk ejection efficiency $\xi$ (always $<0.1$), which are nearly the same for all $\alpha_m$.


The turbulence level parameter $\alpha_m$ has a rather monotonous effect on several disk quantities, regardless of the amount of magnetic turbulent pressure. Figure~\ref{fig:alphap_alpham_disk} shows that as $\alpha_m$ increases, the deviation from the Keplerian rotation rate decreases (i.e., $\delta$ increases). This is because increasing the magnetic diffusivity leads to straighter poloidal and toroidal field lines. As a consequence, the radial magnetic tension becomes less pronounced, and $\delta$ increases (see Eq.~\ref{deltaeq}). For the same reason, the disk is less vertically pinched by the laminar magnetic pressure gradient (Eq.~\ref{eq:P}), leading to more diffusive disks already becoming thicker. This effect is naturally maintained as $\alpha_P$ increases and the disk becomes puffier ($h_d = z_{id}/h$ increases, central right panel in Fig.~\ref{fig:alphap_alpham_disk}). Finally, since both laminar (jets) and turbulent (viscous) torques scale as $\mu^{1/2}$, one expects $\Lambda \propto 1/\alpha_m$, so that increasing $\alpha_m$ increases the effect of the turbulent torque, as observed indeed. The decrease in $\Lambda$ is then amplified as $\alpha_P$ increases. Thus, the stronger the turbulence (large $\alpha_m$ and $\alpha_P$), the less energy and angular momentum the jets extract from the JED.
 

How the accretion sonic Mach number $m_s$ behaves with turbulence is less straightforward. For $\alpha_m=0.5$ and  $\alpha_m=1$, the behavior is similar and monotonous: a larger $\alpha_m$ leads to a more diffusive disk and a larger $m_s$. This can be interpreted as the fact that it is easier to diffuse through the vertical field lines so that the accretion speed can be higher. As $\alpha_P$ increases, leading to a decrease in the toroidal magnetic field at the disk surface and thus a decrease of the jet torque, $m_s$ decreases. But for a rather large $\alpha_m=3$, it is quite surprising to get a {\it smaller} $m_s$ at $\alpha_P=0$ (red curve, Fig.~\ref{fig:alphap_alpham_disk}). This shows that the field lines inside the disk become too straight so the toroidal field is mostly generated only at the disk surface. As a consequence, accretion is lowered at the disk midplane (lack of torque) despite a very large diffusion. The portion where $m_s$ increases with $\alpha_P$ is due to the increase of the disk magnetization $\mu$, which boosts the torques. Eventually, this is overcome by the reducing effect of the magnetic turbulent pressure on the toroidal field, leading to a decrease of $m_s$ with  $\alpha_P$.

For high magnetic diffusivity and turbulent magnetic pressure, no solution can be found because the ejection of matter is so delayed that the resistive MHD region expands until the SM point penetrates the disk. This configuration cannot provide any solution because the magnetization of the disk is no longer constrained.


\begin{figure}
      \resizebox{\hsize}{!}{\includegraphics{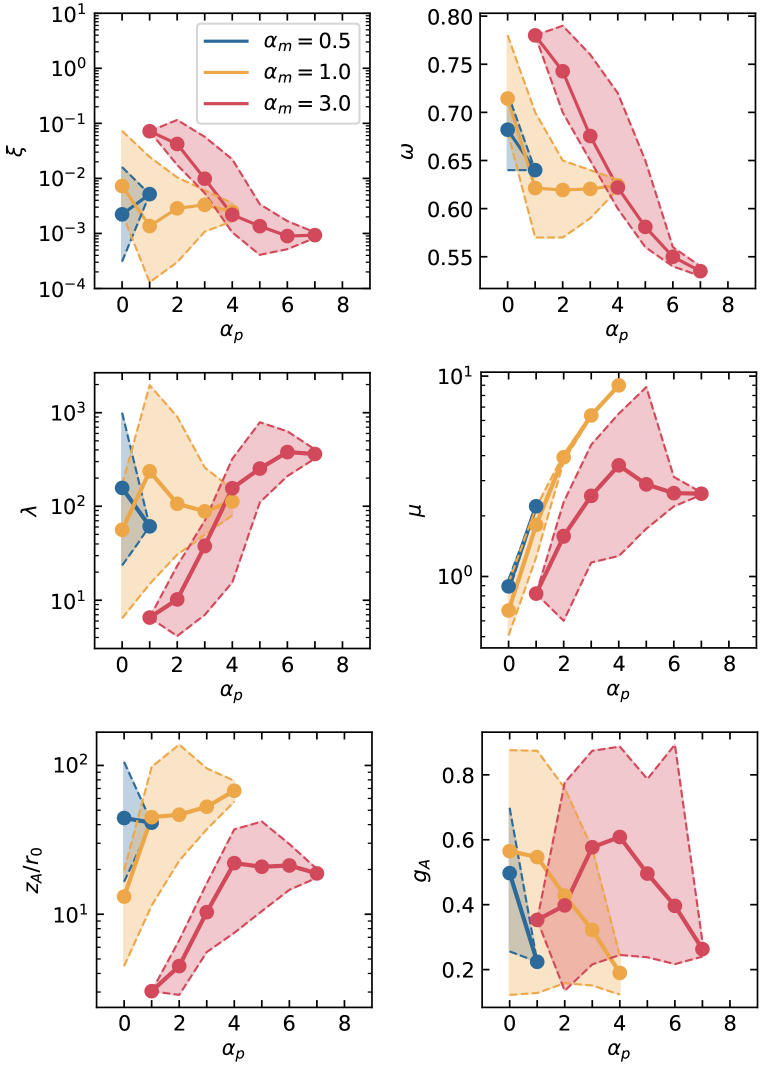}}
      \caption{Influence of $\alpha_P$ on the super-A JED parameter space for various turbulence level $\alpha_m$ obtained with ($\epsilon = 0.1, {\cal P}_m = \chi_m = 1$): disk ejection efficiency $\xi$, rotation rate $\omega$ of the magnetic surface, magnetic lever arm $\lambda$, disk magnetization $\mu$, altitude $z_A/r_o$ of the Alfv\'en point, and the electric current $g_A$ still available at that point. Colored areas correspond to regions accessible by super-A solutions for different magnetic diffusivity levels: $\alpha_m = 0.5$ (blue), $\alpha_m = 1$ (yellow), $\alpha_m = 3$ (red). The corresponding curves represent the average solution for a given $\alpha_P$.}
      \label{fig:alphap_alpham_jet}
\end{figure}



Finally, Figure~\ref{fig:alphap_alpham_jet} presents the impact of $\alpha_m$ on the JED response to an increase of $\alpha_P$, taking into account both SM and Alfv\'enic constraints. As seen in the SM-parameter space, the influence of $\alpha_m$ is quite similar for $\alpha_m = 0.5$ and  $\alpha_m = 1$. 

Although super-A solutions can be found up to $\alpha_P=4$ for $\alpha_m=1$ and only  $\alpha_P=1$ for $\alpha_m=0.5$, the typical ejection index remains qualitatively similar (a few $10^{-3}$ or mass load $\kappa\sim 10^{-3}$, leading to a typical magnetic lever arm $\lambda \sim 10^2$). The reason why super-A solutions cannot be found with increasing $\alpha_P$ is that because the magnetic field curvature at the disk surface becomes too small to allow cold ejection: the ratio $B_r/B_z$ at the SM point decreases below unity. As $\alpha_P$ increases, the disk magnetization $\mu$ increases accordingly, the rotation rate $\omega$ of the magnetic surface decreases (following the decrease of the midplane rotation $\delta$), and the altitude of the Alfv\'en point $z_A/r_o$ increases (best seen with $\alpha_m=1$). All these behaviors are consistent with the increasing disk magnetization and the increasing dominant effect of the poloidal magnetic tension on the jet sub-A region dynamics. This leads to jets that do not propagate much further than the Alfv\'en point, as illustrated by the decrease seen in $g_A$ (bottom right,  Fig.~\ref{fig:alphap_alpham_jet}).


As already seen with the SM constraint, the behavior is profoundly different for large $\alpha_m = 3$. The first surprising result is that no super-A cold solution has been found for $\alpha_P = 0$. The reason is that at such high magnetic diffusivity, the field lines in the disk are too straight to allow cold ejection. However, as $\alpha_P$ increases, the generation of electric currents at the disk surface layers is enhanced, leading to an increase in the ratio $B_r/B_z$ and thus to the existence of super-A outflows. Thus, without turbulent magnetic pressure, no jet would be produced here. Quite interestingly, and in agreement with the SM constraint, the disk magnetization $\mu$ increases less with $\alpha_P$, allowing to achieve super-A outflows closer to the disk (smaller altitude $z_A/r_o$, of course increasing as $\mu$ increases) and potentially propagating much further out from the Aflv\'en point ($g_A$ increasing and greater than 1/2). However, for $\alpha_P>4$, $\omega$ becomes too small, and producing super-A jets requires to use of more and more magnetic energy, leading to a decrease in $g_A$ until there is no super-A solution.

\section{Discussion}

\subsection{Summary and caveats}

We have studied the effect of including a turbulent magnetic pressure in a strongly magnetized accretion-ejection structure (a JED). It turns out that this extra term, which appears only in the vertical and radial momentum equations, has a tremendous effect on the JED parameter space and the cold (negligible enthalpy) jet properties.

\begin{figure*}
    \centering
    \resizebox{0.75\hsize}{!}{\includegraphics{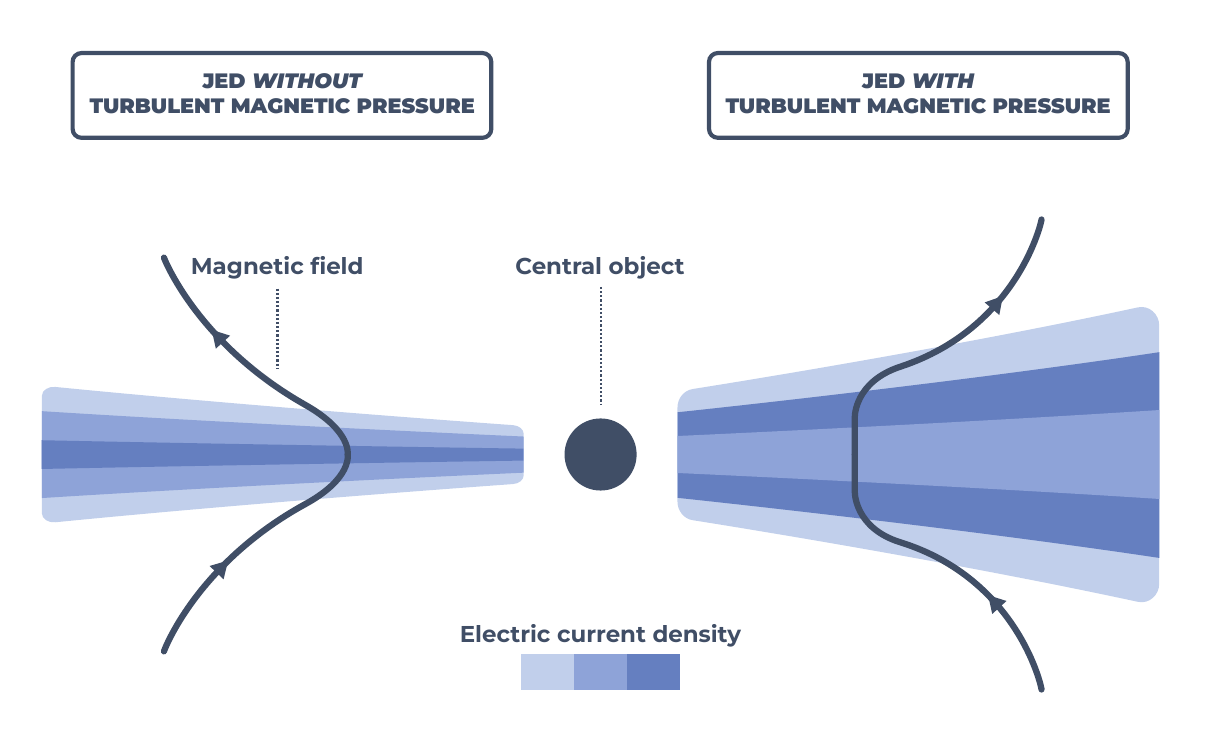}}
    \caption{Schematic representation of the effect of the turbulent magnetic pressure. The turbulent magnetic pressure makes the JED puffier and less electrically conducting, pushing both radial and toroidal electric currents to flow at the disk surface. Field lines become more straight within the disk and their bending as well as their shear occur at the surface. Asymmetry between the two bipolar jets is much easier to achieve because the two poloidal currents can be decoupled within the disk.} 
    \label{fig:crobar}
\end{figure*}

Qualitatively, JEDs with a turbulent magnetic pressure become puffier and appear much less electrically conductive. They tend to force both radial ($J_r$) and toroidal ($J_\phi$) electric current densities to flow at the disk surface, leading to much more straight magnetic (laminar) field configurations within the disk, the bending required for energetically allowing cold jets to be launched occurring mostly at the disk surface (see Fig.~\ref{fig:crobar}). As a consequence, the laminar toroidal magnetic field is also smaller and developed mostly in the upper layers of the disk. This has several consequences. While the midplane accretion remains supersonic, the jet torque is lowered within the disk and the accretion speed is faster in the disk upper layers. Overall, jets from these new solutions carry away less accretion energy and disk angular momentum and also become more collimated due to a stronger poloidal magnetic tension acting in the sub-A region.

There are however two important caveats that need to be discussed, both intimately related to MHD turbulence. The first one is our choice of the vertical profiles of the anomalous (turbulent) quantities. For the sake of simplicity, but also due to the lack of thorough analyzes in the literature, we choose to use simple Gaussian profiles decreasing at a fixed scale. Such Gaussian profiles are not realized in 3D global MHD simulations when the disk is weakly magnetized \citep{zhu18,jacqT21, jacq21}. This numerical situation corresponds to WED solutions found at magnetizations below $\mu \sim 0.1$, where spatial oscillations (MRI channel modes) have been found \citep{jacq19}. A Gaussian profile, or a seemingly decreasing profile on a disk scale height, has been however observed in simulations at large magnetization with $\mu \sim 1$ \citep{jacqT21}. This is the main reason why we restricted our analysis to strongly magnetized JEDs. Our finding that $\alpha_P$ leads to a systematic shrinking of the parameter space may be a consequence of the fact that we do not adjust the vertical scale of the anomalous transport coefficients, that is, take into account the the feedback of the mean flow on the turbulence. Indeed, the prescribed thermal disk scale height $h$, used as the vertical scale of the turbulence, is no longer a valid proxy for the real disk scale height $h_d > h$. This can lead to solutions that did not become super-SM or super-A because they ejected matter too early inside the disk. Taking this new disk scale height into account could be achieved through an iterative method. However, this approach would introduce additional numerical complexities, while the actual profiles of turbulent diffusivities are close to, but not exactly, Gaussian. We have decided to postpone this analysis to future work, where we will use profiles educated by 3D numerical simulations.

A second important aspect that severely limits our parameter space is the simplifying assumption of isothermal (cold) outflows. As $\alpha_P$ increases, the tension associated with the vertical magnetic field becomes increasingly strong, leading to sub-Keplerian accretion disks. This dramatic decrease of the centrifugal drive (and the larger $\epsilon = h/r$, the smaller $\omega$), due to the field lines becoming more and more straight at the disk surface, forbids steady-state ejection to take place. Since \citet{blan82} it is well known that enough poloidal field curvature (bending) is a requirement only for purely magnetically driven outflows, that is, outflows with a negligible enthalpy. \citet{cass00b} have shown that any additional heat input at the disk surface layers deeply modifies the disk vertical structure, allowing for instance to enhance the disk mass loss (they provided a warm super-A solution with $\xi \sim 0.5$). Such surface heating may naturally arise from irradiation coming from a hot central source, which could be either the innermost disk regions or for instance the accretion shock onto the hard surface of the central object. But, as these authors point out, it may also be due to turbulence itself. Now, it is interesting to see that recent 3D numerical GRMHD simulations do display signs of an anomalous dissipation above the disk (see Fig.~12 in \citealt{scep24}). It is still unclear how such a local dissipation, which is withdrawn from the numerical domain at each time step, actually affects the outflow dynamics.

Clearly, MHD turbulence in the disk does more than just provide anomalous transport coefficients and a turbulent magnetic pressure. Another example of a physical process neglected here is turbulent mass diffusion, which could greatly enhance $\xi$. Note that a numerical (due to the grid) mass diffusion can nevertheless occur in nonturbulent simulations, possibly leading to discrepancies between theoretical expectations and numerical experiments (see Sect.~3.5 in \citealt{murp10}). However, addressing all these points is clearly beyond the scope of the present paper.

\subsection{Implications}

Putting aside the previous limitations, let us now discuss our results more quantitatively along with their implications. For our fiducial parameter set ($\epsilon = 0.1, \alpha_m = {\cal P}_m = \chi_m = 1$), classical (i.e., $\alpha_P = 0$) isothermal JED solutions typically have $\xi \sim 10^{-2}$, $\mu \sim 0.5$ and propagate much beyond the Alfv\'en point. When turbulent pressure is included up to its maximum value $\alpha_P = 4$, the ejection efficiency $\xi$ decreases around $10^{-3}$, $\mu$ increases up to $10$, the disk is significantly sub-Keplerian ($\delta \sim 0.6$), jets carry only $\sim \%25$ of the released accretion power and recollimate immediately after crossing the Alfv\'en point. It is therefore highly doubtful that these solutions could be representative of powerful magnetically dominated astrophysical jets propagating much further away from the source. But dealing with the fate of recollimation jets cannot be done within the framework of self-similarity (see for instance \citealt{jann23}).

However, regardless of the limitations discussed previously, the turbulent pressure is highly dependent on the disk aspect ratio $\epsilon$. Specifically, thin disk solutions with $\epsilon = 0.01$ (and $\alpha_m = {\cal P}_m = \chi_m = 1$) are achieved even up to $\alpha_P = 10$, the disk remains nearly Keplerian with $\mu\sim 1$ and jets propagate much further out than thicker disks. They do however have a very low ejection efficiency $\xi \sim 10^{-3}$. Additionally, the turbulent pressure allows for new solutions with high diffusivity level $\alpha_m$ that were not achievable before. For instance, solutions with $\alpha_m = 3$ (and $\epsilon = 0.1, {\cal P}_m = \chi_m = 1$) are sub-Keplerian with $\mu \sim 2$ and still propagate far beyond the Alfv\'en point with $\xi \sim 10^{-2}$.

It therefore appears that obtaining jets that propagate far beyond the Alfv\'en point can be achieved by playing with $\epsilon$ and/or $\alpha_m$ for JEDs at $\mu$ near equipartition with an important turbulent pressure. It turns out that increasing $\alpha_m$ is the best way to enhance the jet mass load, that is, increasing $\xi$ or decreasing the magnetic lever arm $\lambda$. More importantly, classical JED solutions ($\alpha_P=0$) with $\alpha_m=3$ could not be found (in agreement with \citealt{jacq19}). In other words, the presence of a magnetic turbulent pressure ($\alpha_P\neq 0$) appears necessary to obtain more massive, steady-state super-A cold jets at large $\alpha_m$. The physical reason behind this efficient collaboration between $\alpha_m$ and $\alpha_P$ lies in the accretion-ejection geometry depicted in Fig.~\ref{fig:crobar}.

Global 3D MHD simulations of highly magnetized jets, referred to as MAD in the literature, do provide some hints that such a geometry is indeed observed in the numerical experiments. The question now is how to determine in which theoretical parameter space those simulations ended up.  
In fact, the vast majority of numerical papers do not provide the value of the disk magnetization $\mu$. Instead, the beta plasma is usually given but it barely allows to recover our control parameter $\mu$. Indeed the beta plasma is most of the times the ratio of disk averages (not only averages done at the midplane) and, more importantly, it always includes the turbulent magnetic pressure. 

To our knowledge, the only GRMHD paper allowing to infer $\mu, \alpha_P$ and $\epsilon$ is the work of \citet{scep24}. The values provided are measured at $7r_g$ so quite close to the black hole, where general relativistic effects might be already at work. Nevertheless we indicate them as they provide some insight. For the three simulations, the parameter sets ($\epsilon, \mu, \alpha_P$) are: $(0.03, 0.070, 15.1)$, $(0.1, 0.045, 4.7)$ and $(0.3, 0.015, 5.7)$. It can be seen from these values that $\alpha_P$ seems to vary with the disk aspect ratio $\epsilon$, being larger in the thinner disk. It is also worth noting that the JED solutions considered in this paper are actually much more magnetized. This is an indication that we should seek for solutions with large $\alpha_m$ values (or relax the non-oscillatory condition). Unfortunately, there is no measure done in the numerical simulations of the $\alpha_m$ parameter. If we assume that scalings derived from shearing box simulations remain representative of global simulations, then one should seek for $\alpha_m {\cal P}_m \sim 8$ (see Eq.~\ref{eq:salv}). Looking for new solutions of that kind that could best reproduce MAD simulations (including the vertical profiles and outflow behavior) is postponed for future work.

Another long-standing discrepancy between theoretical models and numerical experiments is the value of the disk ejection efficiency. Commonly measured values for $\xi$ varied from 0.5 to 1 (see e.g., \citealt{McKi12}), while in more recent simulations \citet{mani24} found $\xi=0.4$ for example. \citet{scep24} measure values that vary greatly whether the wind is cooled down or allowed to stay warm (due to turbulent heat dissipation beyond the disk surface). For warm winds, $\xi$ goes from 0.12 for $\epsilon=0.03$ to $\xi=1.07$ for  $\epsilon=0.3$, while for cooled winds, $\xi\sim 0.3-07$ for this range in $\epsilon$. Larger disk ejection efficiencies for warm outflows are consistent with theoretical expectations \citep{cass00a} and our results here show that the allowed values for $\xi$ may indeed depend on the disk aspect ratio.

In agreement with the disk global energy budget established between an inner radius $r_{in}$ and an outer radius $r_{out}$, the released accretion power, defined as
\begin{equation}
P_{acc} = \left [ \frac{GM\dot M_a(r)}{2r} \right ]^{r_{in}}_{r_{out}} = \frac{GM\dot M_a(r_{in})}{2r_{in}} \left ( 1 - \left ( \frac{r_{out}}{r_{in}} \right )^{1-\xi} \right ) 
\end{equation}
is positive only if $\xi <1$. Any value of $\xi$ larger than unity shows that either the disk is not yet in steady-state, or that some extra energy is being transferred at $r_{in}$ from the central object. For instance, super-SM solutions with $\xi$ up to 10 are perfectly possible (see Fig.~\ref{fig:alphap_epsilon_disk}), but their Bernoulli invariant $e$ is negative and they cannot become super-A. In numerical simulations, that would correspond to unsteady, very massive bursts of matter that fall back to the disk at larger radii.

To ensure that simulations have indeed reached a steady-state, one should seek for the consistency between the disk ejection efficiency $\xi$ derived from the radial profile of $\dot M_a(r)$ and other disk quantities, such as those in Eq.~\ref{eq:exposants}. It is for instance somewhat troublesome that simulations in \citet{scep24} display a magnetic flux $a \propto r^{3/4}$ (see their Fig.~4) regardless of the disk aspect ratio $\epsilon$, which in theory would necessarily require $\xi<<1$. Note however that these measurements are done quite close to the black hole (up to $10r_g$), where deviations from the Newtonian radial profiles are obviously expected. 

Another possibility, related to turbulence, that might explain the very large disk ejection efficiencies reported in global 3D simulations is the presence of a turbulent mass diffusion. Addressing this last point may however be particularly tricky in numerical simulations. In any case, deeper comparisons between 3D numerical experiments and theory must be carried out.

Note that the discrepancy in $\xi$ also holds for 2D MHD simulations of "alpha" disks, namely simulations where anomalous transport coefficients have been used to mimic turbulence \citep{cass02,zann07,tzef09,murp10,step16}. No clear explanation has been proposed yet, except for a possible numerical effect (see discussion in \citealt{murp10}).

\section{Conclusion}

In this paper, we have presented a major upgrade on the theory of highly magnetized (near equipartition) accretion-ejection structures, also called JEDs. We analyzed the effect of turbulent magnetic pressure on the disk structure and its impact on the overall interrelationship between the disk and its jets. Although this pressure appears only in the radial and vertical momentum equations, it plays a major role and deeply affects the JED parameter space.

The turbulent magnetic pressure makes the disk puffier and less electrically conductive, forcing both radial and toroidal electric currents to flow at the disk surface. Field lines become more straight inside the disk, and their bending and shearing occur mostly at the surface. Accretion remains supersonic but the speed increases at the disk upper layers. For the usual values of the turbulence parameters explored so far in classical (i.e., no turbulent magnetic pressure $\alpha_P=0$) JED models, the inclusion of turbulent pressure leads to a dramatic decrease in the cold jet power and angular momentum extraction from the disk.

However, recent 3D global simulations tend to show that MHD turbulence  provides simultaneously large anomalous magnetic diffusivities and viscosity as well as a large magnetic turbulent pressure. When combined in the semi-analytical model, these large values allow for the existence of new super-Alfv\'enic jet solutions that were previously impossible to obtain.

The next obvious step is to better characterize the vertical profiles involved in MHD turbulence. They may indeed differ from the Gaussian profiles used in this work, which focuses on strongly magnetized disks ($\mu \sim 1$). Also associated with these profiles, some heat dissipation may be always present, providing an additional enthalpy reservoir for the outflows and allowing, for instance, more mass to be ejected. Taking into account non-Gaussian profiles is an absolute necessity in the case of WEDs (with $\mu << 1$). This is postponed for future work. 

Our mathematical description of accretion-ejection structures allows us to incorporate both laminar and turbulent effects. However, it is now necessary to use more accurate values and more realistic vertical profiles for the turbulence, based on the results of converged 3D numerical simulations.

The JED and WED concepts have emerged on the theoretical side, while on the computational side MAD and SANE  configurations have received much attention. A direct comparison between these costly 3D numerical experiments and steady-state theory has finally become feasible.

Our results demonstrate that JEDs provide a state-of-the-art mathematical description of their MAD numerical counterparts. However, further efforts from both sides are needed to firmly establish this point.

\begin{acknowledgements}
    The authors thank Nicolas Scepi for sharing his results and interesting discussions and the anonymous referee for the very insightful comments that have improved the quality of the paper. NZ and JF acknowledge financial support from the CNES French space agency and PNHE, PNPS programs of French CNRS. JJ acknowledges support by the NSF AST-2009884, NASA80NSSC21K1746 and NASA XMM-Newton 80NSSC22K0799 grants.
\end{acknowledgements}

%
%

\bibliographystyle{aa}
\bibliography{bibliography}

\end{document}